\documentclass[namedreferences]{solarphysics}
\usepackage[hyperref,optionalrh,solaromanenum]{spr-sola-addons} 
\usepackage{graphicx}                    

\usepackage[hyperref,optionalrh,solaromanenum]{spr-sola-addons} 
\usepackage{graphicx}                    
\usepackage{lscape}
\usepackage{sidecap}
\usepackage{color}                       
\usepackage{breakurl}                         

\chardef\us=`\_

\begin{document}
\begin{article}
\begin{opening}
\title{Long-term Variations in Solar Activity: Predictions for
Amplitude and  North--South Asymmetry of Solar Cycle 25} 

\author{\fnm{J.}~\lnm{Javaraiah}\orcid{0000-0003-0021-1230}}

\institute{Bikasipura, Bengaluru-560 111,  India\\
Formerly worked at Indian Institute of Astrophysics, 
Bengaluru-560 034, India.\\
corref, email: \url{jajj55@yahoo.co.in}; 
 email:\url{jdotjavaraiah@gmail.com}; email: \url{jj@iiap.res.in}\\}

\runningauthor{J. Javaraiah}
\runningtitle{Prediction  of North--south Asymmetry in Solar Activity}

\begin{abstract}
In most of the solar cycles,  activity in the
 northern and southern hemispheres peaks at different times. 
One hemisphere peaks well before the other and 
 at least one of the hemispheric maxima frequently does not coincide with
 the whole sphere maximum. Prediction for the maximum of a  hemisphere
and the corresponding  north--south asymmetry  of a solar cycle  
 may help  to understand the mechanism of the solar cycle,
 the solar-terrestrial relationship, and solar-activity
 influences on space weather.
 Here we analysed the sunspot-group data from the Greenwich Photoheliographic 
Results (GPR) during 1874\,--\,1976 and Debrecen Photoheliographic Data (DPD) 
during 1977\,--\,2017 and studied the cycle-to-cycle variations in the values
 of 13-month smoothed monthly mean  sunspot-group area in the whole 
sphere (WSGA), northern hemisphere (NSGA), and southern hemisphere (SSGA) at 
the epochs of maxima of Sunspot Cycles 12\,--\,24 and at the epochs of
 maxima of WSGA, NSGA, and SSGA  Cycles 12\,--\,24 (note that solar-cycle 
variation of a parameter is expressed as a cycle of that parameter).
 The cosine fits to the values of   WSGA, NSGA, and
 SSGA at the maxima of sunspot,  WSGA, NSGA, 
and SSGA Cycles 12\,--\,24, and to the values of the corresponding 
north--south 
asymmetry,  suggest the existence  of a $\approx$132-year periodicity in the
activity of the northern hemisphere, a 54\,--\,66-year periodicity  
in the activity of the southern hemisphere, and a 50\,--\,66 year 
periodicity in the north--south asymmetry in activity 
at all the aforementioned epochs. 
  By extrapolating  the  best-fit cosine curves   we 
predicted the amplitudes  and the corresponding north--south asymmetry 
of the 25th  WSGA, NSGA, and SSGA cycles. 
 We find that  on  average Solar Cycle~25   in sunspot-group area    
would be to some extent  smaller than Solar Cycle~24 in sunspot-group area. 
  However, by 
inputting  the predicted  amplitudes  of the  25th  WSGA, NSGA, and
 SSGA cycles  in the linear relationship  
between  sunspot-group area and sunspot number we find that the amplitude 
 ($130\pm12$) of Sunspot Cycle~25 would be  slightly larger than that 
of  reasonably small Sunspot Cycle~24.  Still it confirms that
 the beginning  of the upcoming Gleissberg cycle
  would take place around  Solar Cycle~25.
 We also find that   except at the maximum of 
 NSGA Cycle~25 where the strength of activity in the northern hemisphere
 would be dominant, the strength of activity in the southern hemisphere 
 would be dominant  at the maximum epochs of the 25th sunspot, 
 WSGA, and SSGA cycles.
\end{abstract}

\end{opening}

\section{Introduction}
The existence of  differences between the strengths of 
solar activity in the Sun's northern and southern hemispheres is 
well known  as the north--south asymmetry in solar-activity. 
It exists in most  solar activity phenomena  (\opencite{hath15}).
 Besides this, it is also known that  the epochs of 
maxima of solar cycles   are not the same in the northern and 
southern hemispheres, that is in some solar cycles  the maximum in 
one of the hemispheres does not  coincide  with the maximum of the total 
 (whole sphere/visible whole disk) activity (e.g. see  
Figure~1 in \opencite{jj20} and also see  \opencite{mci13}). 
In some solar cycles the activity peak occurs first in the
 northern hemisphere and in some other cycles, it first occurs  in the southern 
hemisphere.  The existence of  some systematic phase difference between 
 the  cycles of activity  in the northern and southern hemispheres is 
also known (e.g. \opencite{zolo10}; \opencite{ng10}; \opencite {mura12}; 
 \opencite{mci13}; \opencite{rcj21}). There exist $\approx$12-year, 
$\approx$55-year, and $\approx$100-year  periodicities in 
 north--south asymmetry (\opencite{cob93}; \opencite{verma93}; 
 \opencite{jg97}; \opencite{li02}; \opencite{deng16};
 \opencite{jj20}; \opencite{non21}).
 However, the  origin of these as well
 as the known short-term periodicities in the north--south asymmetry
 (e.g. \opencite{knaack04}; \opencite{rj15}; \opencite{chow16}; 
 \opencite{mb16}; \opencite{rcj21}) is not yet clear 
(\opencite{nort14}; \opencite{sc18}; \opencite{nep19}).
Our recent analysis indicates that the $\approx$12-year  and  $\approx$51-year
 periodicities of north--south asymmetry are manifestations of 
the differences in strengths of the corresponding 
periodicities in northern and southern hemispheres and their origin 
 might be connected to the configurations of giant planets (\opencite{jj20}). 
It has been observed  that in several solar 
cycles  there exist some differences in the  maximum epochs of sunspot number
 and sunspots area cycles (\opencite{hath15}). 
 Recently, we have analysed the combined sunspot-group data from
 Greenwich Photoheliographic Results (GPR) during 1874\,--\,1976 and 
 Debrecen Photoheliographic Data (DPD) during 1977\,--\,2017 and 
 have predicted the  north--south asymmetry in the average sunspot-group
 area at 
the maximum epoch of  Sunspot Cycle~25 (\opencite{jj21}).  Here we  analyse
 the aforementioned sunspot-group data and  predict the north--south 
asymmetry  at the maximum epochs of solar cycles  in the northern and 
southern hemispheres. This 
 may help in better understanding the solar dynamo  
and the solar-terrestrial  relationship. Here our approach is 
different from that of \cite{jj21}, where  
we have used the same  method that was used  in our earlier 
articles (\opencite{jj07}; \citeyear{jj08}; \citeyear{jj15}). 
In those earlier articles the prediction was made 
 based on the existence of a good correlation between the amplitude of
 a solar cycle 
and the sum of the areas of sunspot groups in the $0^\circ$\,--\,$10^\circ$
 latitude 
interval of the southern hemisphere just after around one year  from 
the maximum of the preceding  cycle  (hereafter area-sum). 
Here  we study the cycle-to-cycle 
 variations in the maxima of Solar Cycles 12\,--\,24 in the average areas 
of sunspot groups in  northern and southern hemispheres, as well as in 
the whole sphere, by determining the best-fit cosine functions  to these data. 
In principle, from this method it is possible to make  predictions for 
several upcoming solar cycles. However, due to the considerable 
uncertainties in the obtained  best-fit  cosine functions, here we make  
 cautious predictions for the amplitude and north--south asymmetry
 of Solar Cycle 25 only.    

In the next section we describe the data analysis.  In Section~3 
we present the results, and in Section~4 we present the conclusions and  
 briefly discuss  them.      

\section{Data Analysis}
Here the data and  the analysis are the same as in  \cite{jj19}. In 
 \cite{jj21} and here we have used the values of the 
 amplitudes ([$R_{\rm M}$]), i.e. 
 the highest value of 13-month smoothed monthly mean sunspot number, and  
the maximum epochs of sunspot cycles given by  
 \cite{pesnell18}. These were determined by \cite{pesnell18}  
 from the time series of  13-month smoothed monthly mean 
values of  Version~2 international sunspot number (SN) available 
 at \textsf{www.sidc.be/silso/datafiles}.
   The updated GPR (1874\,--\,1976) and DPD (1977\,--\,2017) daily 
sunspot-group data 
were downloaded from the website 
\textsf{fenyi.solarobs.unideb.hu/pub/DPD/} (for details see 
 \opencite{gyr11}; \opencite{bara16}; \opencite{gyr17}).
 These data contain, besides the 
heliographic positions and other parameters, the corrected 
whole-spot area [\,msh: millionth of solar hemisphere] of each
 sunspot group for its
each day observation. First we determined the mean area of 
sunspot groups in the Sun's whole sphere (WSGA), northern hemisphere (NSGA), 
and southern hemisphere (SSGA) during each calender month of the years 
1874\,--\,2017,  and then we determined the  13-month smoothed monthly mean
 values and the corresponding  standard errors ($s = {\sigma}/{\sqrt{13}}$, 
where $\sigma$ is the standard deviation). From the time series of 
the  13-month smoothed monthly mean values we obtained the values and 
the corresponding epochs of the following parameters during
 SN, WSGA, NSGA, and SSGA Cycles 12\,--\,24  (note that for brevity and convenience solar cycle variation of a  parameter
 is expressed as a cycle of that parameter).\\
 $ T_{\rm M}$: epoch of SN cycle maximum,\\ 
$T_{\rm W}$: epoch of WSGA cycle maximum,\\ 
$T_{\rm N}$: epoch of NSGA cycle maximum,\\ 
$T_{\rm S}$: epoch of SSGA cycle maximum,\\ 
It should be noted that  usually in a solar cycle these epochs are not in
 an increasing or decreasing order.  However rarely there could be an order,
  and even  all these four epochs of a solar-cycle could be the same.\\
RWA: value of WSGA at $T_{\rm M}$,\\ 
RNA: value of NSGA at $T_{\rm M}$,\\ 
RSA: value of SSGA at $T_{\rm M}$,\\ 
WAM: value of maximum  of a WSGA Cycle,\\
WNA: value of NSGA at $T_{\rm W}$,\\ 
WSA: value of SSGA at $T_{\rm W}$,\\ 
NAM: value of maximum  of an NSGA Cycle,\\
NSA: value of SSGA at $T_{\rm N}$,\\ 
NWA: value of WSGA at $T_{\rm N}$,\\ 
SAM: value of maximum  of an SSGA Cycle,\\
SNA: value of NSGA at $T_{\rm S}$, and \\
SWA: value of WSGA at $T_{\rm S}$. 

 We calculated a best-fit cosine function 
to the cycle-to-cycle modulation in 
each aforementioned parameter and by extrapolating the 
best-fit cosine curves we obtained the values of 
the above parameters for the 25th WSGA, NSGA, and SSGA cycles.    
We also calculated the best-fit cosine function to the cycle-to-cycle
 modulation of relative north--south asymmetry $(N-S)/(N+S)$, 
where $N$ and $S$ represent the values of the parameters that correspond to 
the northern  and southern  hemispheres, respectively,  at the epochs of   
maxima of sunspot, WSGA, NSGA, and SSGA cycle. We  predict north--south 
asymmetry at the corresponding epochs of the  25th cycles.    
In the  calculations of the best-fit cosine functions, 
  weights equal to the corresponding standard error of
 the parameters are used.
 All  the linear least-squares fits were calculated 
by using the Interactive Digital Library (IDL) software {\textsf{FITEXY.PRO}}, 
which  is downloaded from 
 the website \textsf{idlastro.gsfcnasa.gov/ftp/pro/math/}. 
An advantage of using  this software    
is that the  errors in both  abscissa and  ordinate values will be taken 
into account  
 in the calculation of  linear fits by the least-squares method. 

\section{Results and Predictions}
In Table~1, we have given the   13-month smoothed monthly mean values of 
WSGA, NSGA, and SSGA at the epochs  of maxima of SN,  WSGA, NSGA, and 
SSGA of Cycles 12\,--\,14.  Figure~1 shows    
 the values of the parameters  NAM, SAM, WAM, and
 RWA given in this table  versus time (respective epochs).
 The large (small)  error bars in the values of these parameters 
of most of the large (small) solar cycles are due to the strong (weak) 
 fluctuations  in the  monthly mean sunspot-group area around the maximum 
epochs of the  large (small) solar cycles. 
In Table~2, we  give the information obtained from  
this figure and Table~1 about  dominant hemisphere, phase-leading hemisphere,
 and the hemisphere coinciding with the whole sphere. In the same table,  
 the difference between the epochs of $R_{\rm M}$ and WAM is also given. 
As we can see in Figure~1 and Table~2
in some solar cycles North is dominant and in some other cycles South 
is dominant.  In many solar cycles the maximum in
one of the hemispheres does not  coincide with the maximum of the total
(whole sphere) activity.
In some solar cycles the activity peak occurs first in the
 northern hemisphere and in some other cycles, it occurs first 
 in the southern hemisphere.  

\begin{table}
{\tiny
\caption[]{The values [\,msh] of WSGA, NSGA, and SSGA 
at the epochs $T_{\rm M}$,  $T_{\rm W}$,  $T_{\rm N}$, and  $T_{\rm S}$ 
 of the maxima of SN,  WSGA, NSGA, and SSGA cycles ($n =$ 12, 13,...,24),
  respectively,  determined from the
 corresponding time series of 13-month smoothed monthly
 mean areas of the sunspot groups in whole sphere,
 northern hemisphere, and southern hemisphere. The maximum [$R_{\rm M}$] 
values of SN are also given  
 (all the maximum values are indicated with {\it bold}).}
\begin{tabular}{lccccccccc}
\hline
&&\multicolumn{4}{c}{At maximum $R_{\rm M}$ of SN} &&\multicolumn{3}{c}{At maximum WAM of WSGA}\\ 
\cline{3-6}
\cline{8-10}
\noalign{\smallskip}
$n$ &$T_{\rm M}$ &$R_{\rm M}$& RWA &RNA & RSA & $T_{\rm W}$ & WAM &WNA & WSA\\
12&  1883.96&${\bf 124.4\pm 12.5}$&$ 1371\pm 122$&$  414\pm  75$&$  957\pm 105$&  1883.96&${\bf 1371\pm 122}$&$  414\pm  75$&$  957\pm 105$\\
13&  1894.04&${\bf 146.5\pm 10.8}$&$ 1616\pm 110$&$  621\pm  57$&$  995\pm 100$&  1894.04&${\bf 1616\pm 110}$&$  621\pm  57$&$  995\pm 100$\\
14&  1906.12&${\bf 107.1\pm 9.2}$&$ 1044\pm 140$&$  761\pm 139$&$  283\pm  43$&  1905.45&${\bf 1161\pm 158}$&$  745\pm 137$&$  416\pm  90$\\
15&  1917.62&${\bf 175.7\pm 11.8}$&$ 1535\pm 171$&$  829\pm 124$&$  707\pm  89$&  1917.54&${\bf 1554\pm 165}$&$  853\pm 123$&$  701\pm  92$\\
16&  1928.29&${\bf 130.2\pm 10.2}$&$ 1324\pm 123$&$  631\pm  96$&$  693\pm  65$&  1926.29&${\bf 1467\pm 211}$&$  809\pm 181$&$  658\pm  98$\\
17&  1937.29&${\bf 198.6\pm 12.6}$&$ 2120\pm 176$&$ 1309\pm 140$&$  811\pm 108$&  1937.29&${\bf 2120\pm 176}$&$ 1309\pm 140$&$  811\pm 108$\\
18&  1947.37&${\bf 218.7\pm 10.3}$&$ 2641\pm 210$&$ 1051\pm 121$&$ 1590\pm 237$&  1947.37&${\bf 2641\pm 210}$&$ 1051\pm 121$&$ 1590\pm 237$\\
19&  1958.20&${\bf 285.0\pm 11.3}$&$ 3441\pm 208$&$ 1749\pm 164$&$ 1693\pm 144$&  1957.96&${\bf 3480\pm 239}$&$ 1801\pm 153$&$ 1679\pm 166$\\
20&  1968.87&${\bf 156.6\pm 8.4}$&$ 1556\pm  82$&$  951\pm  83$&$  605\pm  46$&  1970.54&${\bf 1628\pm  93}$&$  926\pm  70$&$  701\pm 121$\\
21&  1979.96&${\bf 232.9\pm 10.2}$&$ 2121\pm 162$&$ 1064\pm 139$&$ 1057\pm 142$&  1981.71&${\bf 2338\pm 177}$&$ 1065\pm 134$&$ 1274\pm 193$\\
22&  1989.87&${\bf 212.5\pm 12.7}$&$ 2269\pm 193$&$ 1121\pm 124$&$ 1148\pm 118$&  1989.45&${\bf 2591\pm 179}$&$ 1401\pm 141$&$ 1190\pm 126$\\
23&  2001.87&${\bf 180.3\pm 10.8}$&$ 2157\pm 206$&$ 1073\pm 111$&$ 1084\pm 152$&  2002.20&${\bf 2334\pm 179}$&$  951\pm 114$&$ 1383\pm 141$\\
24&  2014.29&${\bf 116.4\pm 8.2}$&$ 1600\pm 116$&$  420\pm  40$&$ 1180\pm 144$&  2014.45&${\bf 1629\pm 114}$&$  418\pm  40$&$ 1211\pm 143$\\
\noalign{\smallskip}
&&\multicolumn{3}{c}{At maximum NAM of NSGA} &&\multicolumn{3}{c}{At maximum SAM of SSGA}\\ 
\cline{3-5}
\cline{7-9}
\noalign{\smallskip}
 &$T_{\rm N}$ & NWA &NAM & NSA & $T_{\rm S}$ & SWA &SNA & SAM\\
12&  1882.37&$978\pm 145$&${\bf  476\pm  72}$&$  501\pm 116$&  1883.96&$ 1371\pm 122$&$  414\pm  75$&${\bf  957\pm 105}$\\
13&  1894.12&$ 1585\pm 115$&${\bf  649\pm  61}$&$  935\pm 115$&  1893.96&$ 1585\pm 111$&$  577\pm  55$&${\bf 1008\pm  99}$\\
14&  1906.04&$ 1109\pm 143$&${\bf  822\pm 135}$&$  287\pm  45$&  1907.45&$ 1098\pm 129$&$  512\pm  81$&${\bf  585\pm 103}$\\
15&  1917.71&$ 1492\pm 178$&${\bf  855\pm 116}$&$  638\pm  84$&  1917.62&$ 1535\pm 171$&$  829\pm 124$&${\bf  707\pm  89}$\\
16&  1926.29&$ 1467\pm 211$&${\bf  809\pm 181}$&$  658\pm  98$&  1928.21&$ 1338\pm 121$&$  584\pm 102$&${\bf  754\pm  57}$\\
17&  1937.54&$ 2088\pm 195$&${\bf 1374\pm 158}$&$  714\pm 109$&  1938.62&$ 1855\pm 164$&$  715\pm  91$&${\bf 1141\pm 151}$\\
18&  1949.62&$ 2060\pm 196$&${\bf 1200\pm 114}$&$  860\pm 153$&  1947.45&$ 2625\pm 215$&$ 1011\pm 129$&${\bf 1614\pm 232}$\\
19&  1959.54&$ 2847\pm 233$&${\bf 2220\pm 227}$&$  627\pm  77$&  1957.79&$ 3428\pm 257$&$ 1673\pm 183$&${\bf 1755\pm 142}$\\
20&  1967.62&$ 1596\pm 142$&${\bf 1108\pm 100}$&$  488\pm  73$&  1969.87&$ 1549\pm 113$&$  767\pm 103$&${\bf  783\pm 105}$\\
21&  1979.37&$ 2067\pm 164$&${\bf 1211\pm 111}$&$  856\pm 117$&  1981.96&$ 2293\pm 199$&$  976\pm 123$&${\bf 1317\pm 179}$\\
22&  1989.45&$ 2591\pm 179$&${\bf 1401\pm 141}$&$ 1190\pm 126$&  1991.54&$ 2405\pm 167$&$  894\pm 129$&${\bf 1511\pm 189}$\\
23&  2000.96&$ 1831\pm 167$&${\bf 1122\pm 119}$&$  709\pm  65$&  2002.20&$ 2334\pm 179$&$  951\pm 114$&${\bf 1383\pm 141}$\\
24&  2011.71&$  905\pm 116$&${\bf  687\pm 100}$&$  218\pm  40$&  2014.45&$ 1629\pm 114$&$  418\pm  40$&${\bf 1211\pm 143}$\\
\hline
\end{tabular}
\label{table1}
}
\end{table}

\begin{figure}
\centering
\setcounter{figure}{0}
\includegraphics[width=\textwidth]{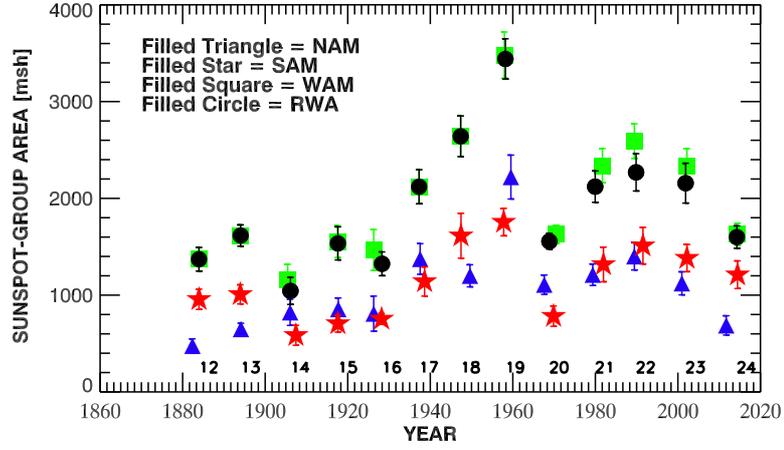}
\setcounter{figure}{0}
\vspace{-1.0cm}
\caption{Values given in Table~1  versus
 time (corresponding epochs) of the parameters:  
NAM, i.e. the values of maxima of NSGA Cycles~12\,--\,24; 
SAM, i.e. the values of maxima of SSGA Cycles~12\,--\,24; 
WAM, i.e. the values of maxima of WSGA Cycles~12\,--\,24; and 
RWA, i.e. the values of the  13-month smoothed monthly mean 
sunspot-group areas at the maxima of Sunspot Cycles~12\,--\,24.} 
\label{f1}
\end{figure}

\begin{table}
\caption[]{The information obtained by using Table~1 and 
 Figure~1 about  dominant hemisphere (DH), phase-leading 
hemisphere (LH), and whether the hemisphere coincides/is close to the  
whole sphere (CH) in a solar cycle~($n$). The  corresponding  differences 
between the values of parameters and  between the epochs  are 
given within parentheses. In the last column the  differences
 between $T_{\rm M}$ and $T_{\rm W}$ are also given. 
In the 2nd column the ``same" implies that there is no phase difference.  
 In the 4th and 5th columns a zero/small absolute value 
 indicates that the corresponding peaks   are coincident/very close.}
\begin{tabular}{lccccc}
\hline
$n$& DH (NAM$-$SAM) &LH ($T_{\rm N}-T_{\rm S}$)&CH 
($T_{\rm N}-T_{\rm W})/(T_{\rm S}-T_{\rm W}$)& $T_{\rm M}-T_{\rm W}$ \\
\hline
12&south ($-481$)&north ($-1.59$)&south ($-1.59$/$0.0$)&$0.0$\\
13&south ($-359$)&same ($0.16$)&both ($0.08/-0.08$)&$0.0$\\
14&north ($237$)&north ($-1.4$)&north ($0.59/2.0$)&$-0.67$\\
15&north ($148$)&same ($0.09$)&both ($0.17/0.08$)&$-0.08$\\
16&north ($55$)&north ($-1.92$)&north ($0.0/1.92$)&$-2.0$\\
17&north ($233$)&north ($-1.08$)&north ($0.25/1.33$)&$0.0$\\
18&south ($-414$)&south ($2.17$)&south ($2.25/0.08$)&$0.0$\\
19&north ($465$)&south ($1.75$)&south ($1.58/-0.17$)&$-0.24$\\
20&north ($325$)&north ($-2.25$)&south ($-2.92/-0.67$)&$1.67$\\
21&south ($-106$)&north ($-2.59$)&south ($-2.34/0.25$)&$1.75$\\
22&south ($-110$)&north ($-2.09$)&north ($0.0/2.09$)&$-0.42$\\
23&south ($-261$)&north ($-1.24$)&south ($-1.24/0.0$)&$0.33$\\
24&south ($-524$)&north ($-2.74$)&south ($-2.74/0.0$)&$0.16$\\
\hline
\end{tabular}
\label{table2}
\end{table}

\begin{figure}
\centering
\setcounter{figure}{0}
\includegraphics[width=8.5cm]{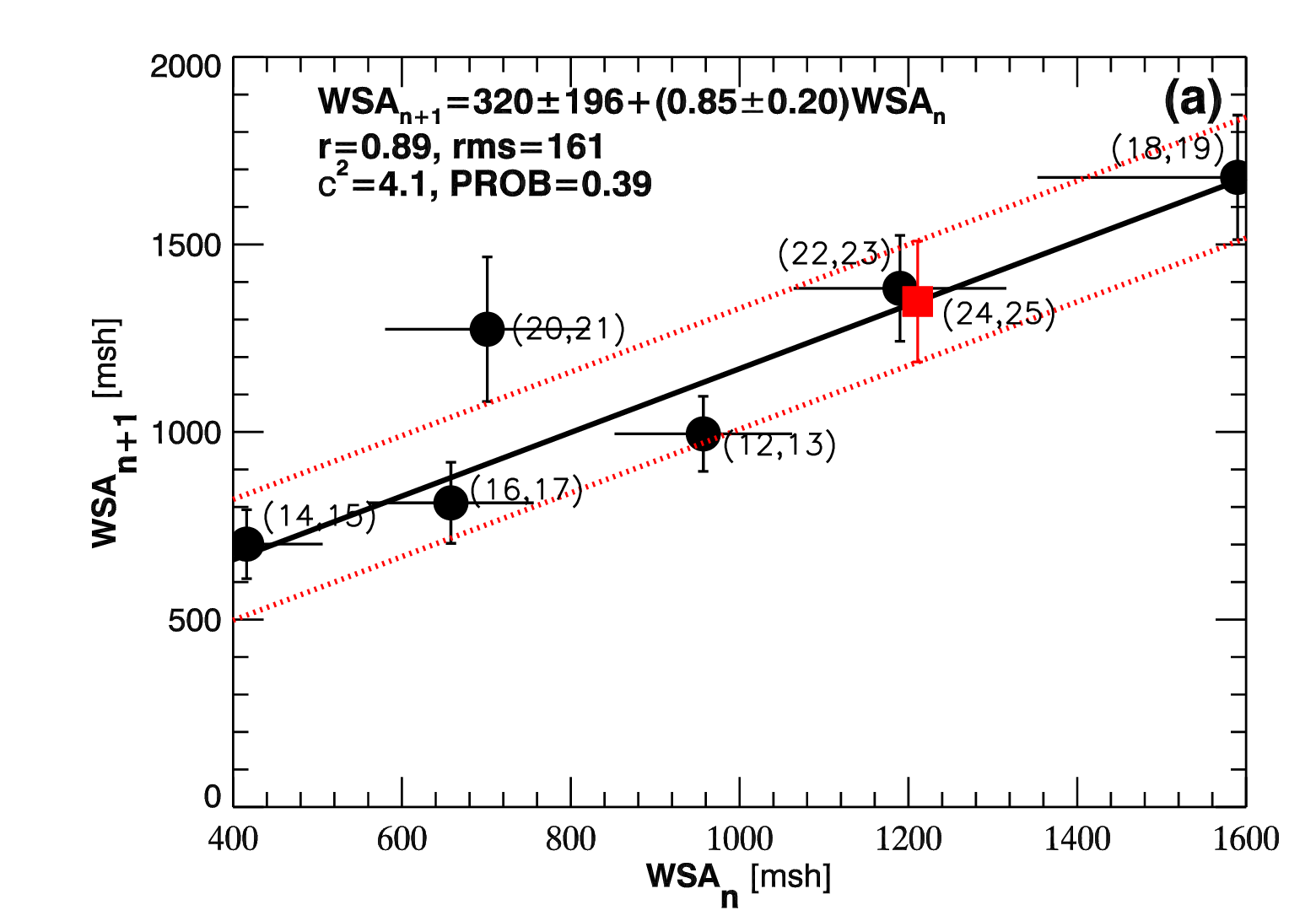}
\includegraphics[width=8.5cm]{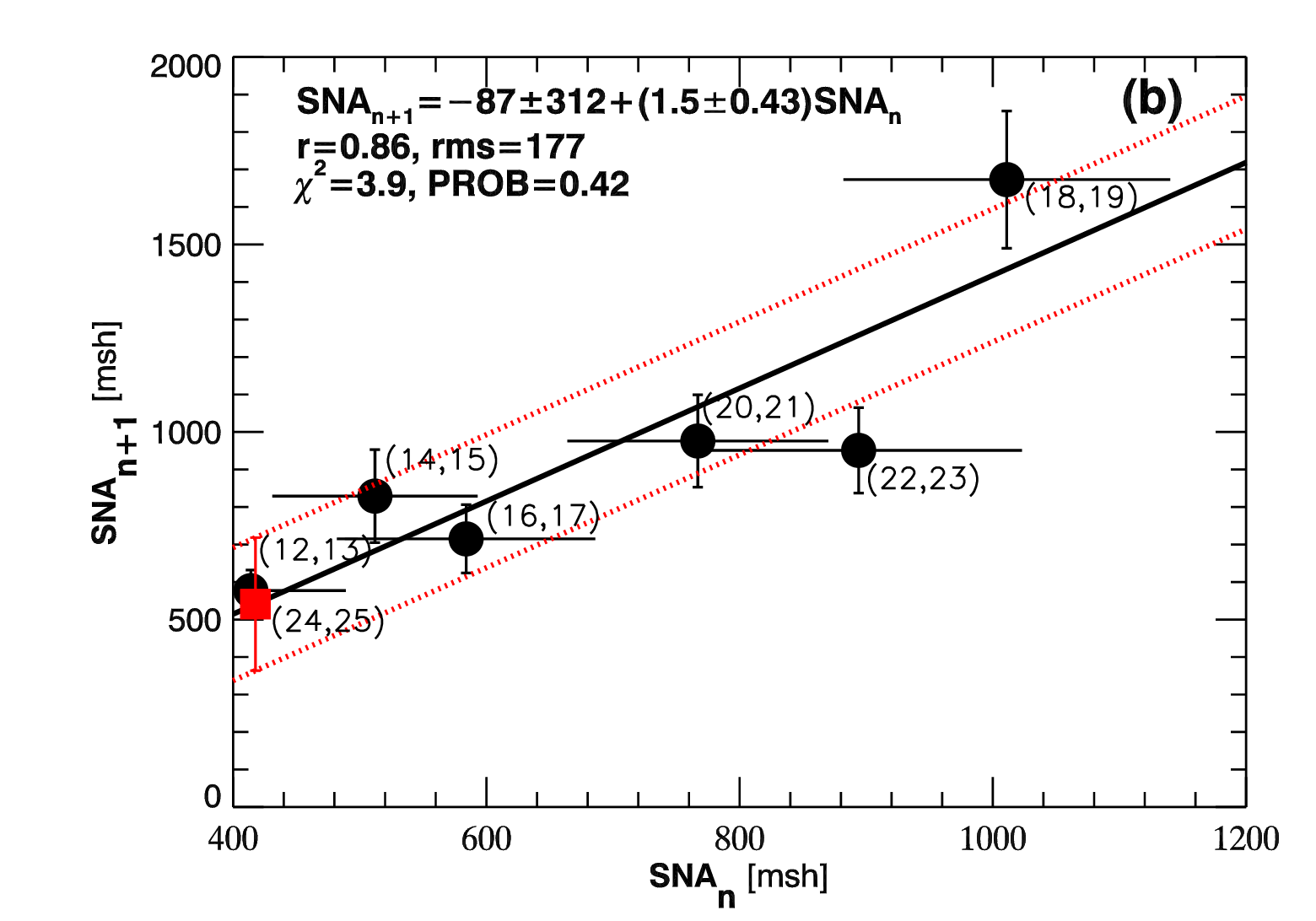}
\setcounter{figure}{1}
\caption{({\bf a})  WSA$_{n}$ of an 
even-numbered WSGA cycle  versus  WSA$_{n+1}$ of  following
odd-numbered WSGA cycle and  ({\bf b})  SNA$_{n}$ of an 
even-numbered SSGA cycle  versus  SNA$_{n+1}$ of  following 
odd-numbered  SSGA cycle. The {\it horizontal and vertical error bars} 
represent  the standard errors of the abscissa and the ordinate, 
respectively; both  are taken into account  in the calculations of  
linear least-square fits.
The  {\it continuous line} represents the  best-fit linear 
relationship. The {\it dotted lines} ({\it red})  are drawn at one-rms 
(root-mean-square 
deviation) levels. The best-fit linear equation and the values of  
 correlation coefficient [$r$], $\chi^2$ and the corresponding 
probability (PROB), and rms  are also shown. The 
{\it filled squares} ({\it red})
 represent the predicted values of WSA and SNA at the epochs of 
maxima of WSGA Cycle~25 and SSGA Cycle~25.} 
\label{f2}
\end{figure}

 In the case of sunspot number, the amplitude of Solar Cycle~21 is larger than  
that of Solar Cycle~22, whereas, as can be seen in Figure~1, in the case of 
sunspot area the behavior  is opposite to that of sunspot number, i.e.
 Solar Cycle~22 is larger than 
Solar Cycle~21  (\opencite{hath15}). This difference could be because 
the ratio of  small to large numbers of  sunspots/sunspot groups  may be 
 larger in Solar Cycle~21 than in Solar Cycle~22. 
 The aforementioned opposite behavior  is 
also found in the cases of  whole sphere,  northern, and southern 
hemispheres' mean areas 
of sunspot groups (i.e. RWA,  RNA, and RSA here) at the epochs of maxima of 
Solar Cycles~21 and 22 (see also \opencite{jj19}). As we can see 
in Table~1,   except in the case of SNA, this property exists 
in the case of
 all remaining  parameters. The value of SNA of SSGA Cycle~21 is larger than
 that of SSGA Cycle~22.      

 According to the Gnevyshev--Ohl rule (G--O: \opencite{go48}), 
 an odd-numbered sunspot cycle is larger than  its preceding even-numbered 
sunspot cycle. However, sunspot-cycle pair (22, 23)   violated this rule, 
i.e. $R_{\rm M}$ of Sunspot Cycle 22 is larger than that of 
Sunspot Cycle~23.  As we can see in Table~1,  WSA values of   WSGA cycle 
pair (22, 23)
  and SNA values of SSGA cycle pair (22, 23) satisfied the G--O rule and 
the corresponding 
 pairs  of each of the remaining all parameters violated  the G--O rule  
  (here the G--O rule in WSA and SNA tentatively 
  means that the cycle-to-cycle  modulations in WSA and SNA  imitate  
 the  G--O rule of solar cycles).

Figure~2  shows the linear relationships between  
 WSA pairs of even- and odd-numbered  WSGA cycles  and between SNA pairs of 
 even- and odd-numbered SSGA cycles. 
 Besides the obtained linear equation, 
the values of $\chi^2$ and the corresponding probabilities (PROB), and rms  
(root-mean-square deviation) are also shown in Figure~2.
The probabilities of the corresponding values of  $\chi^2$  of  the best-fit 
linear relationships are somewhat small. Note that 
 the $\chi^2$-probability is
 a scalar (between zero and unity)  giving the probability that a correct model would give
 a value equal or larger than the  observed $\chi^2$.
A small value of PROB indicates a poor fit, perhaps because  the
 errors are underestimated.

By using the   linear relationships shown in  Figure~2 
we get the values  $1348 \pm 161$ \,msh and $542 \pm 177$ \,msh for  WSA 
of WSGA Cycle-25 and SNA of  SSGA Cycle 25, respectively. These  
 values are larger than the corresponding observed values of 
WSGA Cycle~24 and SSGA Cycle~24  
 implying  satisfaction of the G--O rule.  By using the predicted  value of WSA 
 and the existence of reasonably good correlations  (good linear 
relationships)    
between WSA and SAM,  SAM and SWA, SWA and WAM, WAM and RWA, 
and RWA and $R_{\rm M}$   we  obtained  the values for
  WAM of WSGA Cycle~25  and $R_{\rm M}$ of Sunspot Cycle~25, 
which were found to be much larger than the corresponding values of the 
respective 24th cycles. We have not given them here because 
the aforementioned predicted  values of WSA of WSGA Cycle~25 and SNA of 
SSGA Cycle~25  are to some extent unreliable. It should be noted that
without prior knowledge about  non-violation of the G--O rule by an upcoming
 even- and odd-numbered solar cycle pair, by using the G--O rule  it is not
 possible to predict the amplitude of the odd-numbered solar cycle.

We calculated the cosine fits to the values of the parameters 
of WSGA, NSGA, and SSGA Cycles 12\,--\,24 given in Table~1.  
In Figures 3\,--\,6,  we  show the best-fit cosine curves.
    In Table~3, we  give the results 
obtained by extrapolating the best-fit cosine curves of 
different parameters, i.e. the value of  period, the 
predicted corresponding 25th cycle  value of a  parameter, and 
the values of rms and $\chi^2$. 
The cosine fit of each parameter seems to be  
 reasonably good because only one (mostly the data point of Cycle~19)
 or at most two data points are outliers (away from the one-rms level). 
In principle, we can extrapolate the best-fit cosine curves for 
several  cycles. However, the $\chi^2$ values are large, 
 i.e. the cosine best fits are not very accurate. Therefore the
 corresponding results are only suggestive rather than compelling. Hence 
 we restricted our conclusions to only  the  predictions 
 for the corresponding 25th cycle (although  the
values obtained   for the corresponding  26th  cycle are
 also shown in Figures~3\,--\,6).

We can see in Table~3 that the periodicity in a parameter of  the  
northern hemisphere is  $\approx$132 years, whereas it is  54\,--\,66 years   
in a parameter of the  southern hemisphere, i.e. approximately half of that of 
 the northern hemisphere. This  is consistent with the similar results 
 noticed  in the earlier  analysis (\opencite{jj19}). 
 Morlet-wavelet analysis were also suggested that  a $\approx$51-year 
periodicity in 
the 13-month smoothed area of sunspot groups of the northern hemisphere  is 
 much weaker than that of the southern hemisphere (\opencite{jj20}).

The cosine fits to the data of  RWA, WAM, NWA, and SWA  are 
found to be much more uncertain, and therefore we do not show them.
We get the corresponding 25th cycle  
values of these parameters from the predicted values 
of the parameters given in Table~3.\\ 
${\rm RWA} = {\rm RNA}+{\rm RSA} = 1317\pm123$\,msh,\\
${\rm WAM} = {\rm WNA}+{\rm WSA} = 1402\pm124$\,msh,\\
${\rm NWA} = {\rm NAM}+{\rm NSA} = 1082\pm100$\,msh,\\
${\rm SWA} = {\rm SNA}+{\rm SAM} = 1354\pm116$\,msh. 

The rms  appears large for the 
corresponding  predicted value of a parameter of Solar Cycle~25 given 
Table~3. 
However, the range of values of the parameter is  
large and the predicted 
value is at the minimum level of the corresponding long-term cycle
 (Gleissberg cycle). 
The uncertainties in the  values of  RWA, WAM, NWA, and SWA obtained above are
 reasonably small. They are  determined as
 $\sqrt{({\rm rms}_{\rm_N}^2 + {\rm rms}_{\rm_S}^2)/13}$, 
 where ${\rm rms}_{_{\rm N}}$ and ${\rm rms}_{_{\rm S}}$ are the rms values of 
northern and southern hemispheres' parameters, respectively. 

We also determined the best-fit cosine function to the values of  
$R_{\rm M}$ of SN Cycles~12\,--\,24. It is shown in Figure~7,  
and the details are also given in the last row of Table~3.
 The obtained value for $R_{\rm M}$
 of SN Cycle~25 is slightly larger than that of SN Cycle~24, but the 
former has a large uncertainty (rms value). Moreover,  the cosine fit of
 $R_{\rm M}$ is not good, since the corresponding $\chi^2$ is very large.

\begin{figure}
\setcounter{figure}{1}
\centering
\begin{subfigure}
\setcounter{figure}{2}
\includegraphics[width=5.6cm]{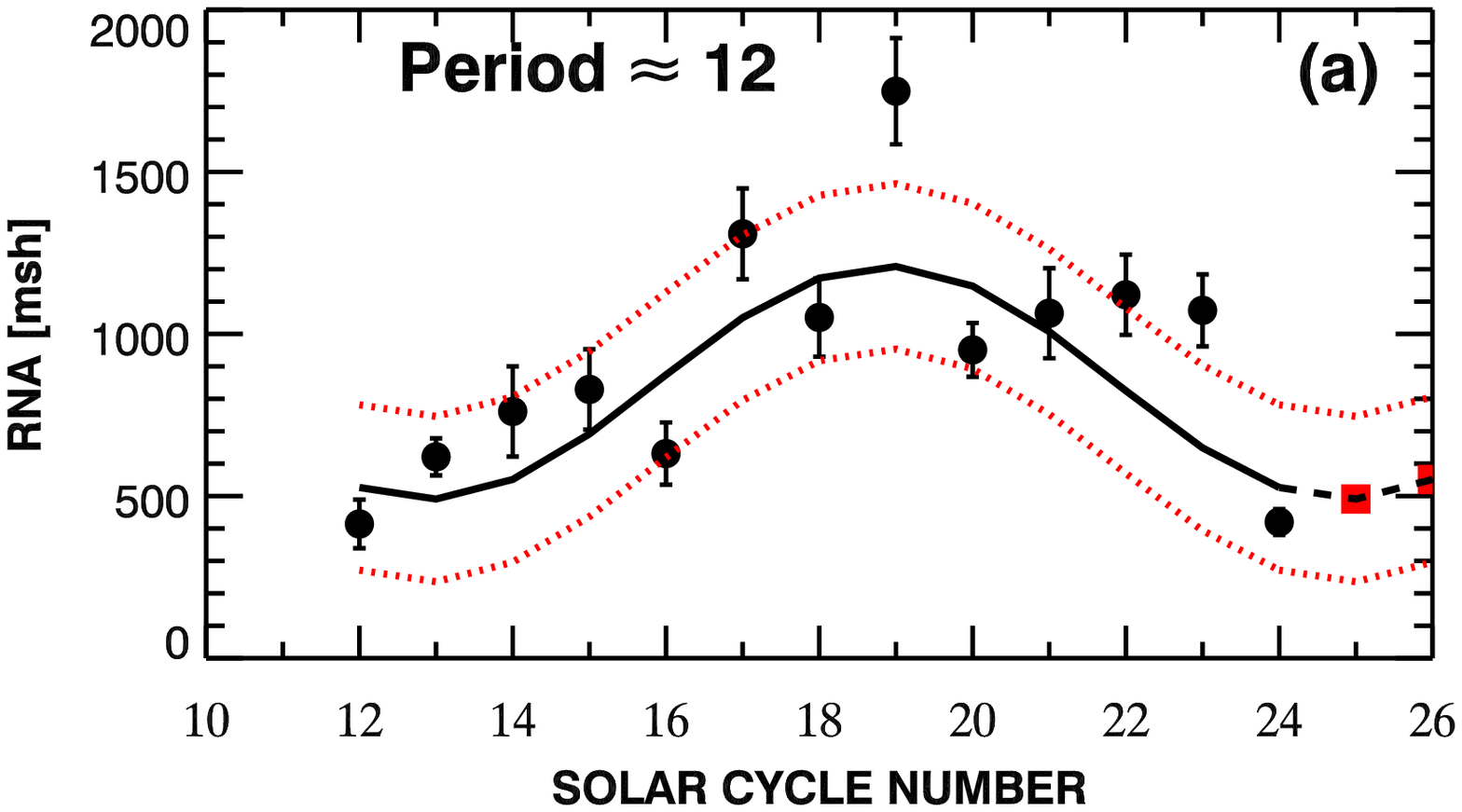}
\includegraphics[width=5.6cm]{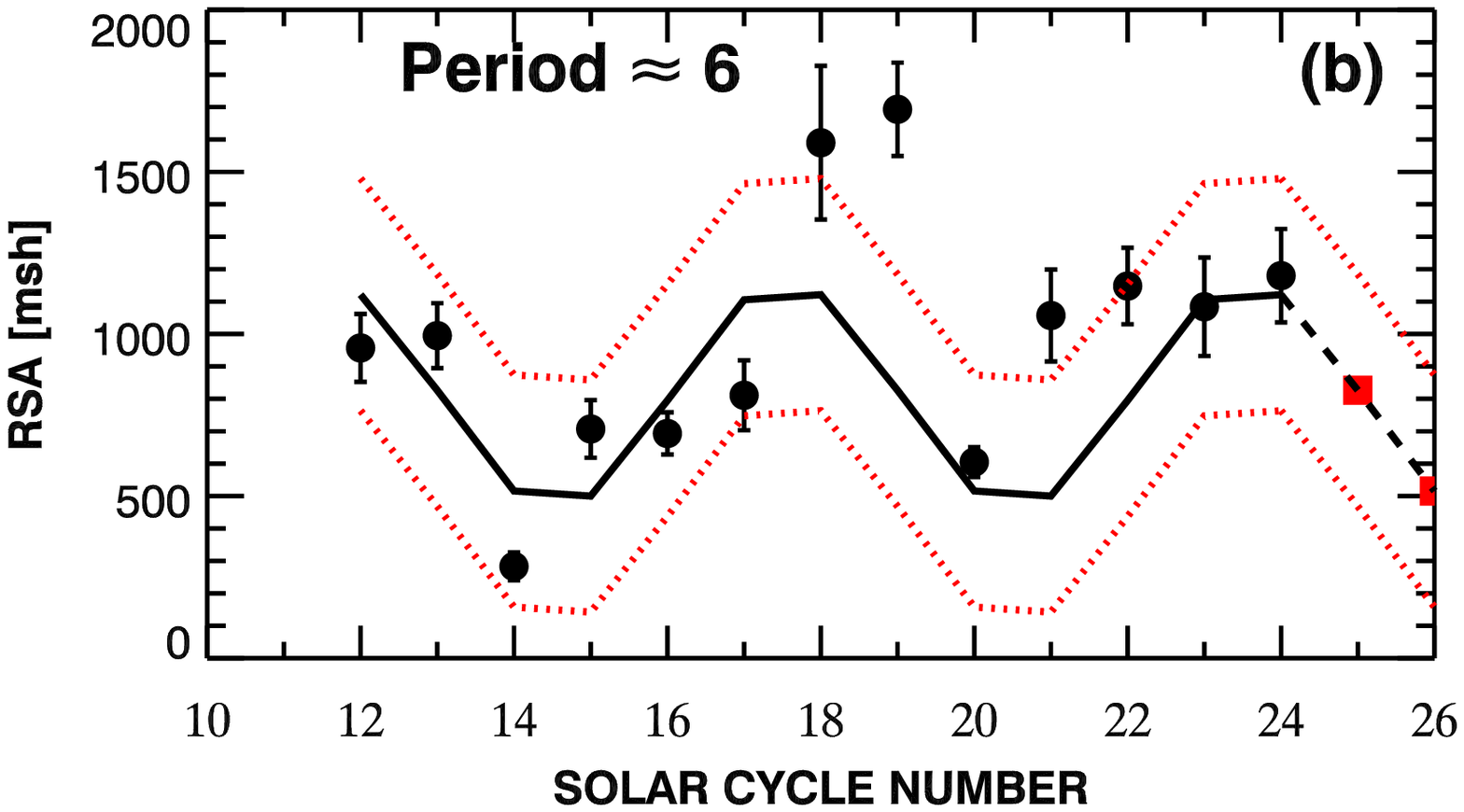}
\end{subfigure}
\caption{{\it Continuous curve} represents the best-fit cosine function
to the values  ({\it filled circles})  ({\bf a}) of RNA and
 ({\bf b}) of RSA   at the epochs of maxima of  Sunspot Cycles 12\,--\,24, i.e 
the values of NSGA and SSGA  at $T_{\rm M}$. 
The  {\it dotted curve} ({\it red})
represents the one-rms  level. The  extrapolated portion is shown as a  
{\it dashed curve} and  the {\it filled squares} ({\it red}) represent
 the predicted values
 of RNA and RSA at the maximum epochs of Sunspot Cycles~25 and 26. 
 The  period (in number of solar cycles)  of the cosine function 
 is also shown. See also the details given in Table~3.} 
\label{f3}
\end{figure}

\begin{figure}
\setcounter{figure}{2}
\centering
\begin{subfigure}
\setcounter{figure}{3}
\includegraphics[width=5.6cm]{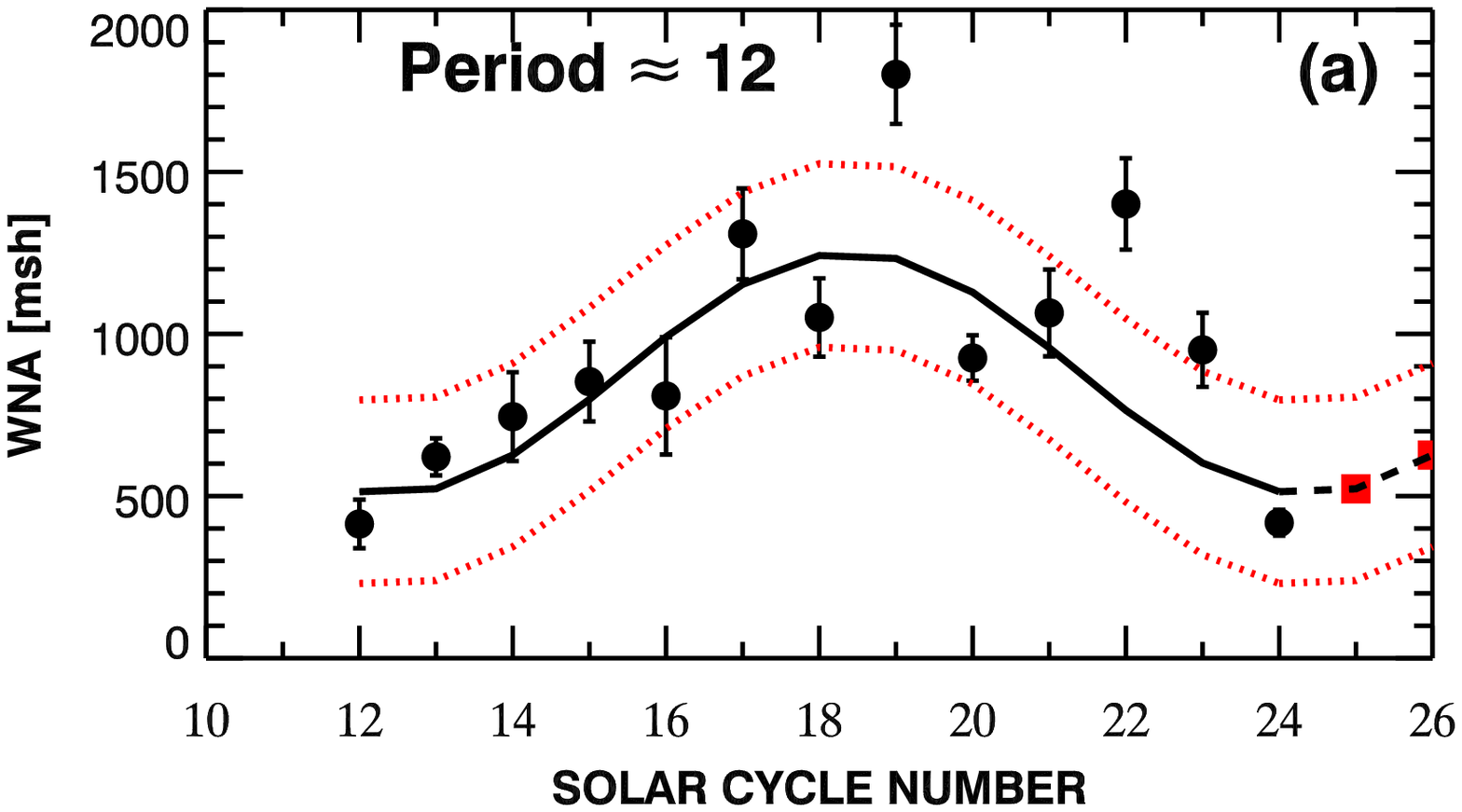}
\includegraphics[width=5.6cm]{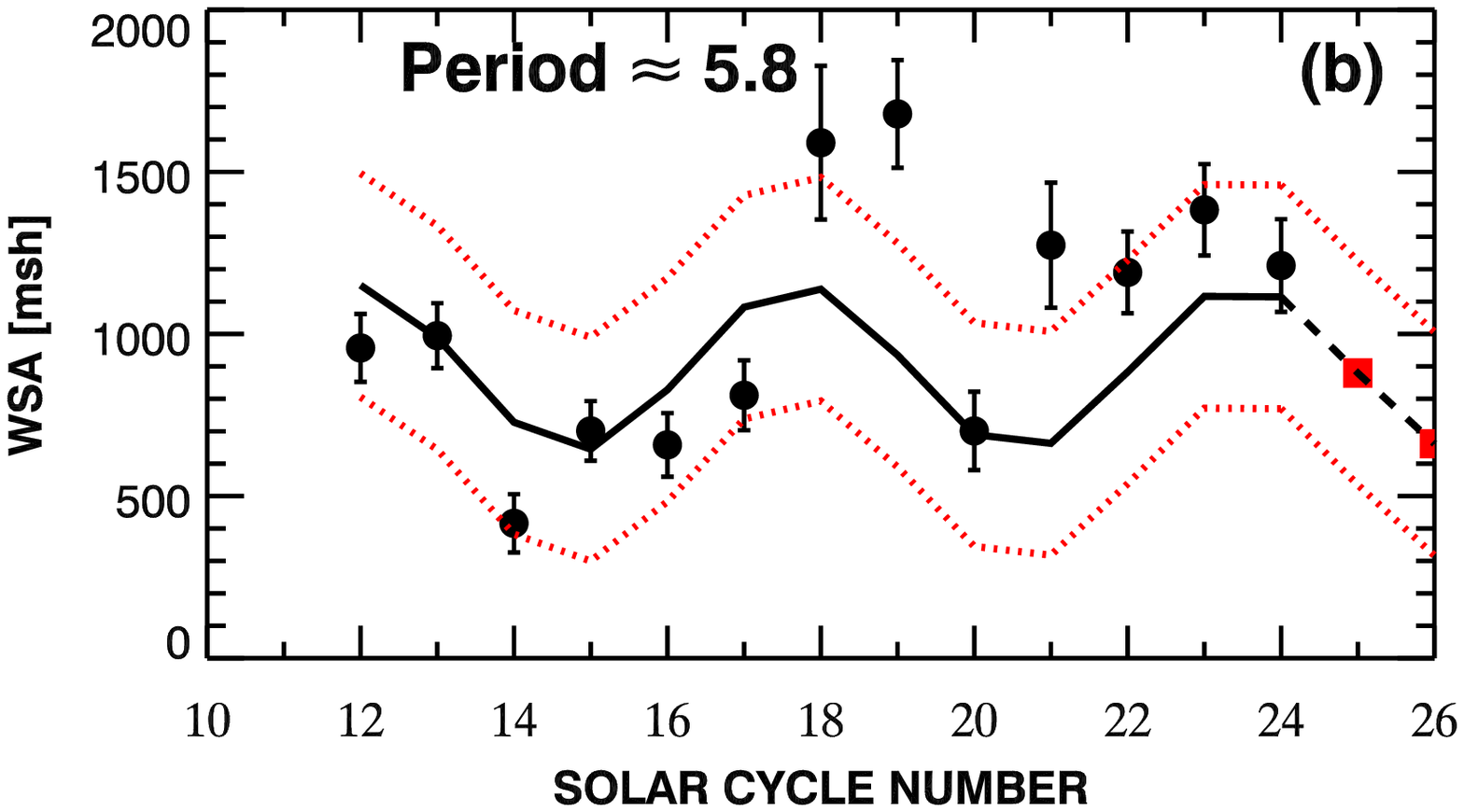}
\end{subfigure}
\caption{{\it Continuous curve} represents the best-fit cosine function
to the values  ({\it filled circles})   ({\bf a}) of WNA and
 ({\bf b}) of WSA of  WSGA Cycles 12\,--\,24, i.e. the values  of
 NSGA and SSGA at  $T_{\rm W}$. The  {\it dotted curve} ({\it red})  
represents the one-rms  level. The  extrapolated portion is shown as a 
{\it dashed curve} and  the {\it filled squares} ({\it red})  represent 
the predicted values
  of WNA and  WSA at the maximum epochs of WSGA Cycles~25 and 26. 
 The  period (in number of solar cycles)  of the cosine function is
 also shown. See also the details given in Table~3.}
\label{f4}
\end{figure}

\begin{figure}
\setcounter{figure}{3}
\centering
\begin{subfigure}
\setcounter{figure}{4}
\includegraphics[width=5.6cm]{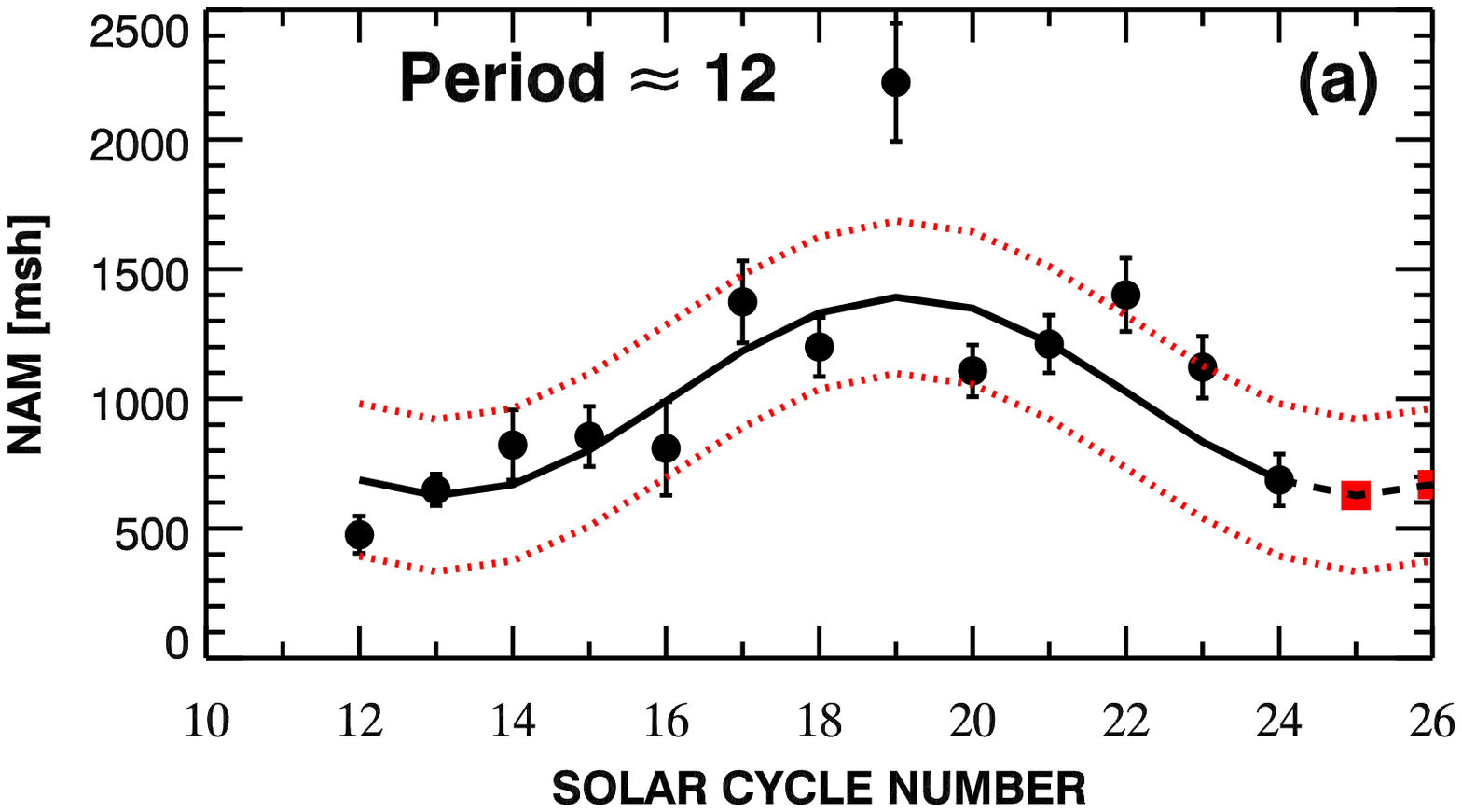}
\includegraphics[width=5.6cm]{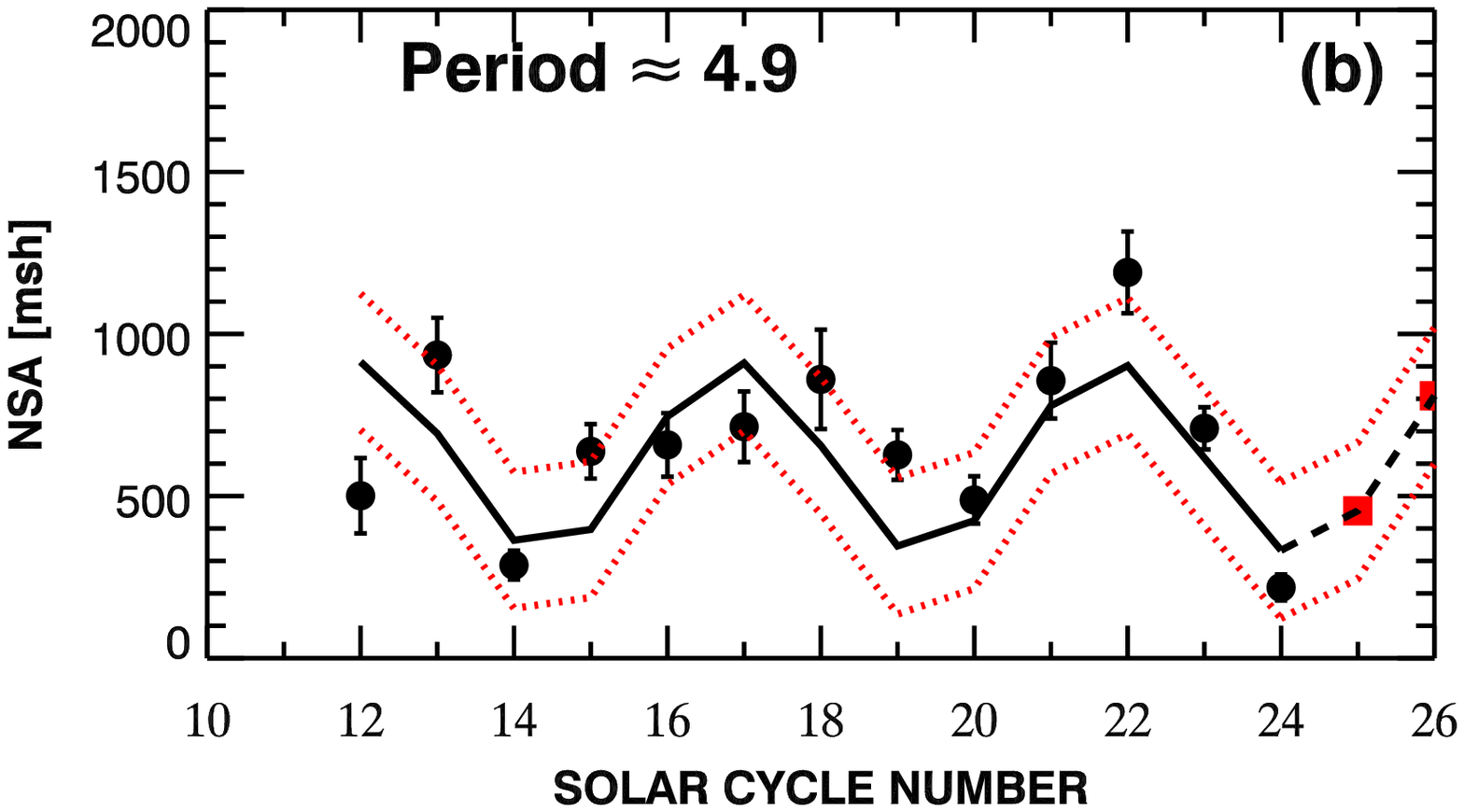}
\end{subfigure}
\caption{{\it Continuous curve} represents the best-fit cosine function
to the values  ({\it filled circles})  ({\bf a})  NAM, i.e. the 
 maximum values of NSGA Cycles 12\,--\,24   and
 ({\bf b}) NSA, i.e. the values of  SSGA  at $T_{\rm N}$.
  The {\it dotted curve} ({\it red})  represents the one-rms  level. The  
extrapolated portion is shown as a {\it dashed curve} and  the 
{\it filled squares} ({\it red}) represent the predicted values of NAM and
NSA at the maximum epochs of NSGA Cycles~25 and 26.
 The  period (in number of solar cycles) of the cosine function 
  is also shown. See also the details given in Table~3.}
\label{f5}
\end{figure}

\begin{figure}
\setcounter{figure}{4}
\centering
\begin{subfigure}
\setcounter{figure}{5}
\includegraphics[width=5.6cm]{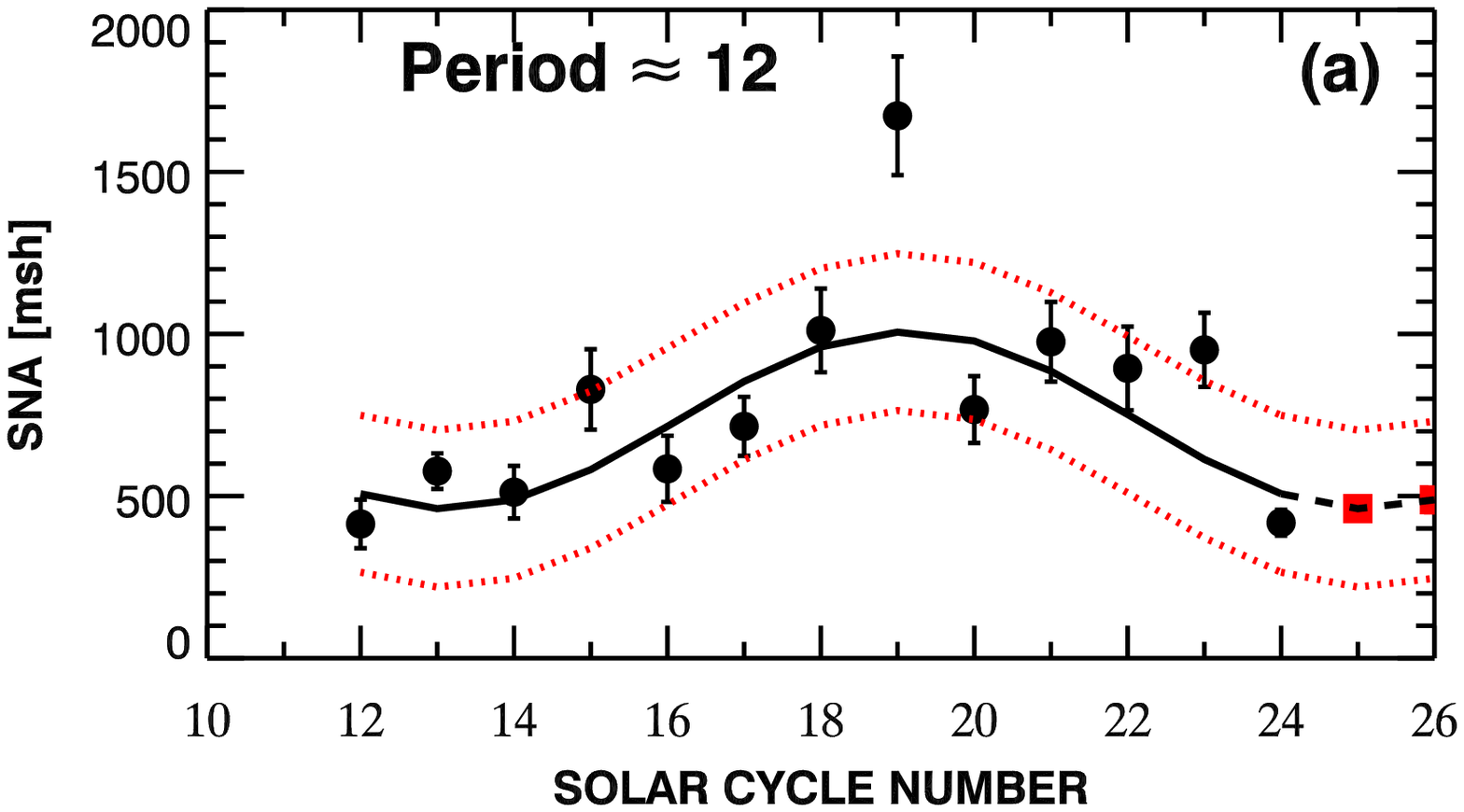}
\includegraphics[width=5.6cm]{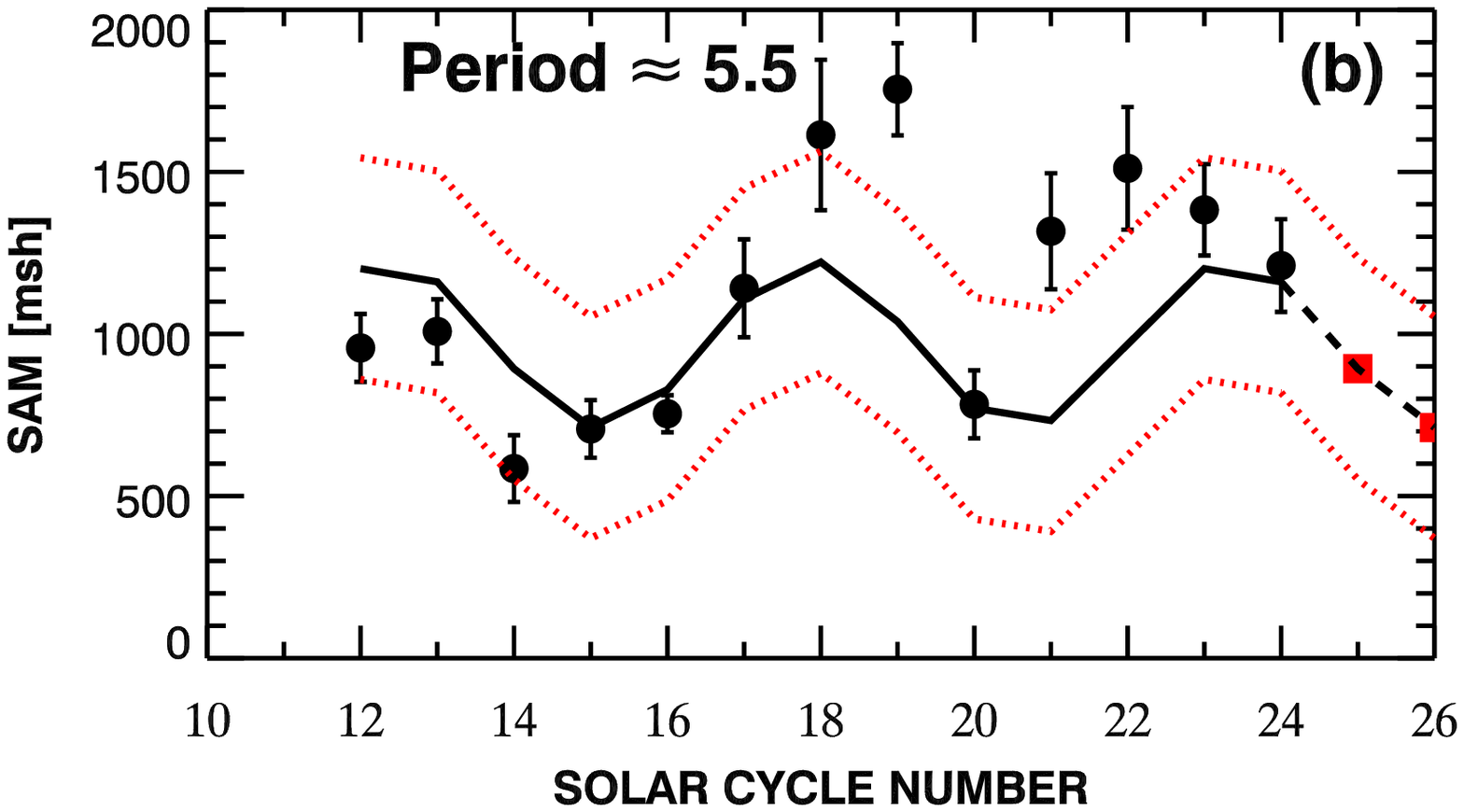}
\end{subfigure}
\caption{{\it Continuous curve} represents the best-fit cosine function
to the values  ({\it filled circles})  ({\bf a})  SNA, i.e. the values 
of  NSGA  at $T_{\rm S}$,   and ({\bf b})  SAM, i.e. the  maximum values 
of SSGA Cycles 12\,--\,24.  The {\it dotted curve} ({\it red}) represents
 the one-rms 
 level. The  extrapolated portion is shown as a {\it dashed curve} and  the 
{\it filled squares} ({\it red}) represent the predicted values of SNA
 and SAM at the maximum epochs of SSGA Cycles~25 and 26. 
 The  period (in number of solar cycles) of the cosine function  is also
 shown. See also the details given  in Table~3.}
\label{f6}
\end{figure}

\begin{SCfigure}
\setcounter{figure}{6}
\centering
\vspace{0.8cm}
\hspace{-0.5cm}
\includegraphics[width=7.0cm]{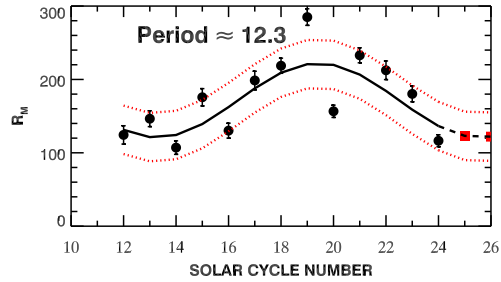}
\caption{\scriptsize {\it Continuous curve} represents the best-fit
 cosine function
to the values  ({\it filled circles}) of $R_{\rm M}$, i.e. the values of
 the amplitudes of Sunspot Cycles 12\,--\,24. 
 The {\it dotted curve} ({\it red}) represents the one-rms  level. 
The  extrapolated portion is shown as a {\it dashed curve} and  the 
{\it filled squares} ({\it red}) represent the predicted values of SN at the
 epochs of maxima of Sunspot Cycles~25 and 26. 
 The  period (in number of solar cycles) of the cosine function is
 also shown. See also the details given  in Table~3.}
\label{f7}
\end{SCfigure}

\begin{SCtable}
\caption[]{\scriptsize The periods of best-fit cosine functions of  
the different parameters given in Table~1  and   
 the parameter  values (\,msh, in the case of a parameter of area) 
  in  Solar Cycle~25 predicted by 
extrapolating the  best-fit cosine curves. The corresponding  values 
of rms and  $\chi^2$ are also given. Note that the average period of  solar
 cycles is $11.03 \pm 1.18$-year (Pesnell, 2018.)}
{\tiny
\begin{tabular}{lccccc}
\hline
Parameter& Period & Period &Pred. Cycle 25 & rms&$\chi^2$\\
         &[cycles] &[years]\\
\hline
RNA&$\approx 12$&$\approx 132$&491&255&66\\ 
RSA&$\approx 6$&$\approx 66$&826&358&118\\ 
WNA&$\approx 12$&$\approx 132$&522&283&69\\ 
WSA&$\approx 5.8$&$\approx 64$&880&345&69\\ 
NAM&$\approx 12$&$\approx 132$&627&294&46\\ 
NSA&$\approx 4.9$&$\approx 54$&455&210&64\\ 
SNA&$\approx 12$&$\approx 132$&461&242&47\\ 
SAM&$\approx 5.5$&$\approx 61$&893&342&67\\ 
$R_{\rm M}$&$\approx 12.3$&$\approx 136$&123&33&141\\
\hline
\end{tabular}
\label{table3}
}
\end{SCtable}

\begin{SCfigure}
\setcounter{figure}{6}
\centering
\setcounter{figure}{7}
\vspace{-0.5cm}
\hspace{-1.0cm}
\includegraphics[width=8cm]{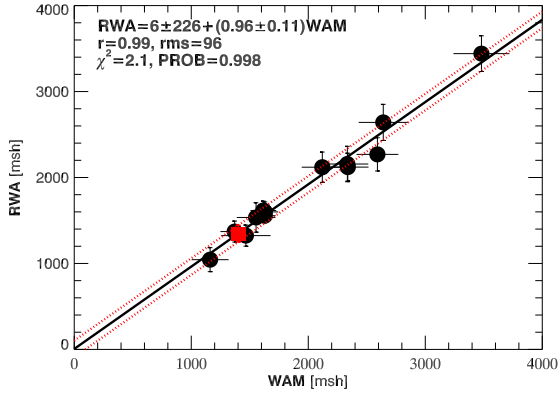}
\caption{\scriptsize Scatter plot of RWA during  Sunspot Cycles 
12\,--\,24 versus WAM  during WSGA Cycles 12\,--\,24. 
The {\it continuous line} represents
 the linear least-square best-fit to the data. 
 The {\it dotted line} ({\it red}) represents the one-rms  level. 
The obtained linear equation and 
 values of the corresponding $r$, rms,  
$\chi^2$,  and PROB  are  given. The {\it filled square} ({\it red})
represents the predicted value of RWA, i.e. the value of WSGA at the 
 epoch of $R_{\rm M}$ of  Solar Cycle~25.}    
\label{f8}
\end{SCfigure}

\begin{SCfigure}
\setcounter{figure}{8}
\hspace{-1.0cm}
\includegraphics[width=8cm]{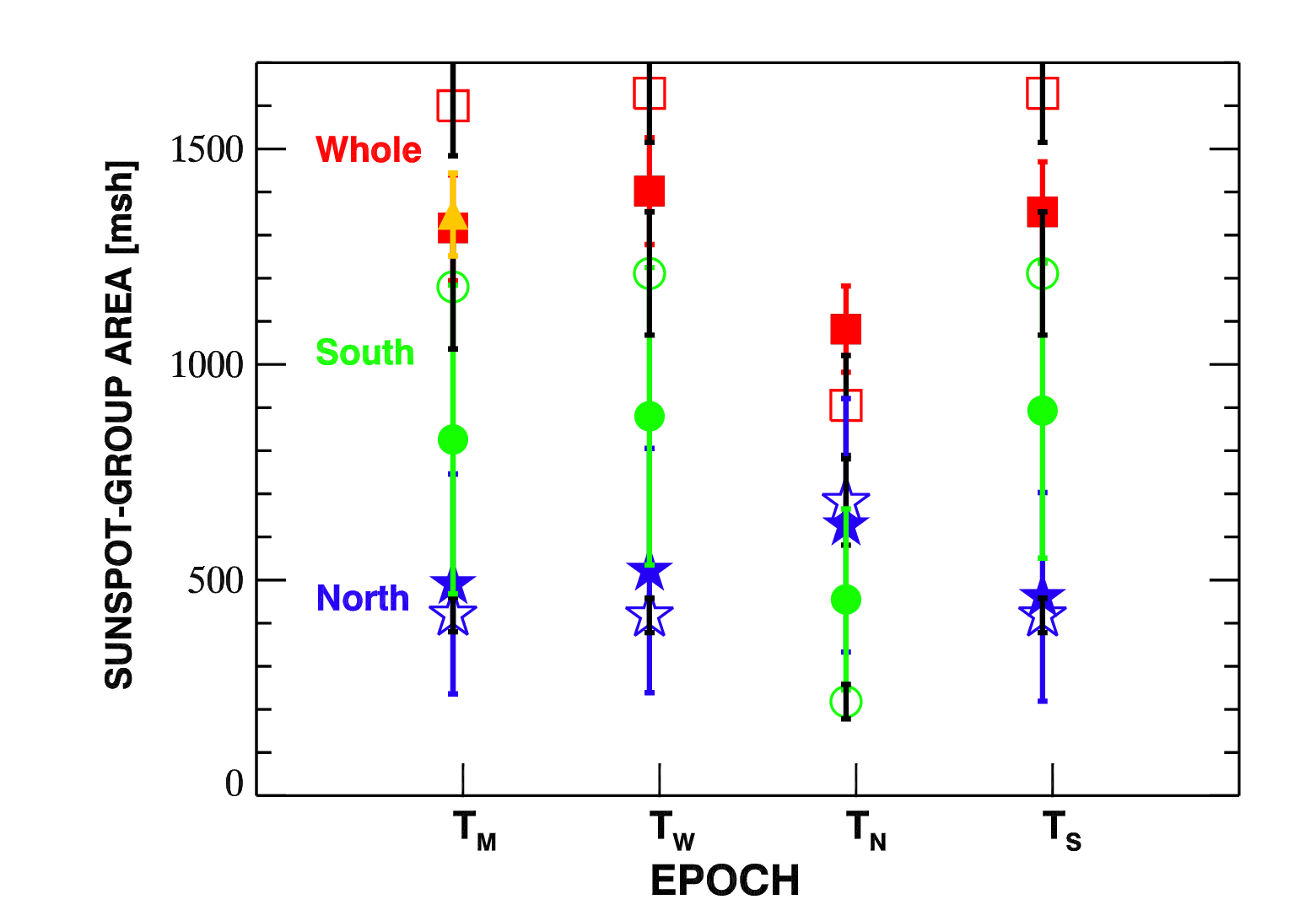}
\caption{\tiny predicted values of the mean sunspot-group area 
   at   $T_{\rm M}$, $T_{\rm W}$, $T_{\rm N}$,
 and $T_{\rm S}$,  i.e. at the epochs of maxima   
 of the 25th  SN, WSGA, NSGA, and SSGA cycles, respectively   
 ({\it filled symbol}) and the corresponding  observed values of  
the 24th cycles ({\it open symbol}) versus the maximum epochs
(the differences between the epochs of $X$-axis are not equal).
The symbols {\it square}, {\it star}, and {\it circle} 
represent the corresponding 
 WSGA, NSGA and SSGA values, respectively.
{\it First column:} RWA, RNA, and RSA at  the epochs of maximum 
  of  SN Cycles~24 and 25,
{\it Second column:}  WAM, WNA, and WSA at the maximum of
 WSGA Cycles~24 and 25,
{\it Third column:} NWA, NAM, and NSA at the maximum of
 NSGA Cycles~24 and 25, and  
{\it Fourth column:} SWA, SNA, and SAM  at the maximum of SSGA
 Cycles~24 and 25. 
 The {\it filled triangle} represents the value of RWA obtained from the 
 WAM--RWA relationship  shown in Figure~8.}
\label{f9}
\end{SCfigure}

 Figure~8 shows the correlation between WAM and RWA.
 Obviously, there  exists a high correlation between 
WAM and RWA. 
The corresponding   linear least-squares  best-fit is highly statistically 
significant ($\chi^2 = 2$, PROB = 0.998).  Using the above predicted value
 of WAM  in WAM--RWA relationship shown in Figure~8,  we get the  value 
$1348\pm96$ \,msh for RWA of SN Cycle~25, which seems to be
 more reliable than the
  value of RWA obtained above by summing the values predicted for RNA 
and RSA (note that the value of $\chi^2$ corresponding to  the  best-fit 
cosine function  of RSA is very high). However,  the two values are almost
 equal.

In Figure~9 we compare the predicted values of the parameters 
at the epochs of the maxima of the 25th  SN, WSGA, NSGA, and SSGA cycles  
 with the observed values of the corresponding  24th cycles.  
As we can see in this figure  there is an indication  
that except at the  peaks of both the NSGA Cycles~24 and 25
 where the activity in the northern hemisphere looks to be negligibly larger
 than that in southern hemisphere, in all the  remaining other
 occasions, including at the epoch of $R_{\rm M}$,  the southern hemisphere is 
dominant. The difference between the predicted values of all 
 northern and southern hemispheres' parameters of the 25th cycles at a given
 epoch is not significant with respect to the large uncertainties in 
these values, suggesting that there exists  no significant north--south 
asymmetry in Solar Cycle~25, whereas most of the corresponding 
differences of the 24th cycle seem to be statistically significant.
The overall pattern of  Solar Cycle~25 
would be closely similar to that of Solar Cycle~24.
The value predicted here for RWA is much larger than
 the value $\approx$701\,msh of RWA (i.e. $A_{\rm W}$) predicted  
by \cite{jj21} by a 
different method, which is much smaller than the RWA of Solar Cycle~24.
The above predicted values of RWA and WSA are reasonably   smaller  than 
the corresponding values of  Solar Cycle~24, and there are
 no significant differences 
in the corresponding values of WNA. The predicted  value of WAM of
 Solar Cycle~25 is  also 
slightly smaller than that of  Solar Cycle~24,  suggesting that 
 on  average  Solar Cycle~25 in sunspot-group area would be smaller
 than  the  corresponding Solar Cycle 24. However, the   
uncertainties  in the predicted values of Solar Cycle~25 are large. Therefore 
all the above results are only suggestive rather than compelling.

\begin{SCfigure}
\setcounter{figure}{8}
\centering
\setcounter{figure}{9}
\vspace{0.3cm}
\hspace{-1.0cm}
\includegraphics[width=8cm]{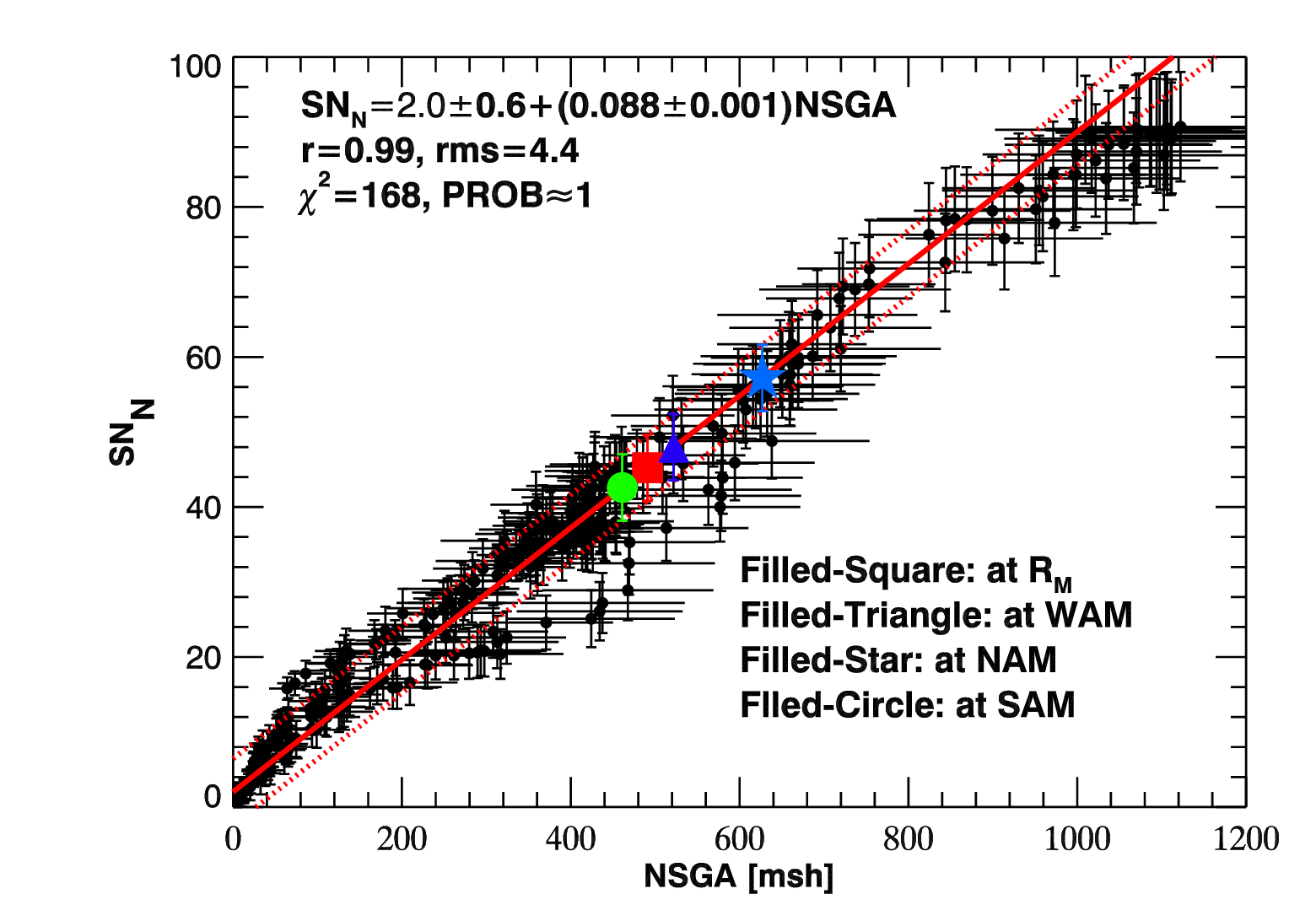}
\caption{\scriptsize Scatter plot of 13-month smoothed monthly mean 
northern hemisphere's sunspot number 
(${\rm SN}_{\rm N}$)  versus NSGA during the period 1992\,--\,2017 
(300 data points). 
 The  {\it continuous line} ({\it red})  represents
 the linear least-squares best-fit to the data. 
 The {\it dotted line} ({\it red}) represents the one-rms  level.
The obtained linear equation and 
the values of the corresponding $r$, rms,  
$\chi^2$, and PROB are  given. The predicted values of 
${\rm SN}_{\rm N}$  at the epochs of $R_{\rm M}$, WAM, NAM, and NSA of  
 the 25th SN, WSGA, NSGA, and SSGA cycles are also shown.}
\label{f10}
\end{SCfigure}

\begin{SCfigure}
\setcounter{figure}{9}
\centering
\setcounter{figure}{10}
\vspace{0.4cm}
\hspace{-1.0cm}
\includegraphics[width=8cm]{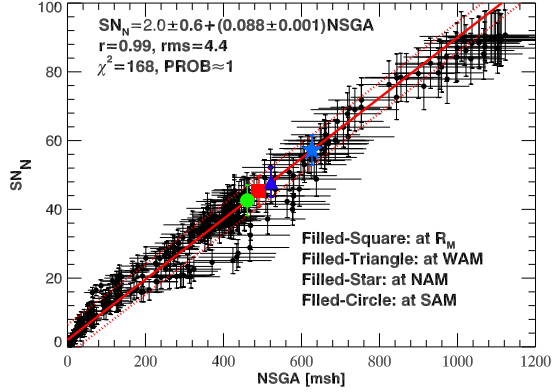}
\caption{\scriptsize Scatter plot of 13-month smoothed monthly mean 
southern hemisphere's sunspot number 
(${\rm SN}_{\rm S}$)  versus SSGA during the period 1992\,--\,2017 
(300 data points). The  {\it continuous line} ({\it red}) represents
 the linear least-squares best fit to the data.
 The {\it dotted line} ({\it red}) represents the one-rms  level.
 The obtained linear 
equation and the values of the corresponding  $r$,
 rms, $\chi^2$, and PROB are  given. The predicted values of 
${\rm SN}_{\rm S}$  at the epochs of $R_{\rm m}$, WAM, SNA, and SAM of  
  25th SN, WSGA, NSGA, and SSGA cycles are also shown.}
\label{f11}
\end{SCfigure}

\begin{SCfigure}
\setcounter{figure}{10}
\centering
\setcounter{figure}{11}
\vspace{1.4cm}
\hspace{-1.0cm}
\includegraphics[width=8cm]{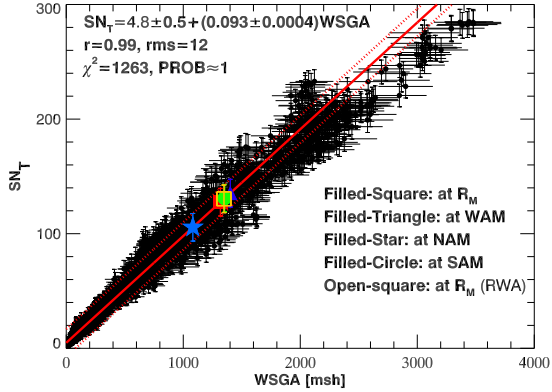}
\caption{\scriptsize Scatter plot of 13-month smoothed monthly mean 
whole-sphere (total) sunspot number 
(${\rm SN}_{\rm T}$)  versus WSGA during the period 1874\,--\,2017 
(1713 data points). The  {\it continuous line} ({\it red}) represents
 the linear least-squares best-fit to the data.
 The {\it dotted line} ({\it red}) represents the one-rms  level.
 The obtained linear equation and 
the values of the corresponding correlation coefficient $r$, rms, 
$\chi^2$, and PROB are  given. The predicted values of 
${\rm SN}_{\rm T}$  at the epochs of $R_{\rm M}$, WAM, NAM, and SAM of  
the 25th SN, WSGA, NSGA, and SSGA cycles are also given.
The value of ${\rm SN}_{\rm T}$ obtained by using RWA predicted from the  
WAM--RWA relationship shown in Figure~8 is also given.}
\label{f12}
\end{SCfigure}

\begin{table}
{\scriptsize
\caption[]{Predictions for  the values of SN in 
 northern hemisphere (${\rm SN}_{\rm N}$),
 southern hemisphere (${\rm SN}_{\rm S}$),
 and whole sphere (${\rm SN}_{\rm T}$) at 
 the epochs  $T_{\rm M}$, $T_{\rm W}$, $T_{\rm N}$, and  $T_{\rm S}$ 
 of the maxima of 25th SN, WSGA, NSGA, and SSGA cycles, respectively,
 by using  in the  linear relationships shown in Figures~10\,--\,12 
the  values of the different  parameters predicted from 
their respective best-fit cosine functions. The symbol $^\mathrm{a}$
indicates that the value of RWA is obtained from WAM--RWA relationship
 shown in Figure~8 (note that ${\rm SN}_{\rm T}$ at $T_{\rm M}$ is the 
same as $R_{\rm M}$).  In the last column of the bottom panel,  
we give the values of ${\rm SN}_{\rm T}$ obtained by 
   summing the corresponding predicted values of 
${\rm SN}_{\rm N}$ ({\it top panel}) and 
 ${\rm SN}_{\rm S}$ ({\it middle panel}).}

\begin{tabular}{lccc}
\hline
\multicolumn{3}{c}{Predictions for the values of ${\rm SN}_{\rm N}$}\\
At the epoch &Predicted value&Used value of\\
$T_{\rm M}$ &$45\pm4.4$&RNA\\
$T_{\rm W}$&$48\pm4.4$&WNA\\
$T_{\rm N}$&$57\pm4.4$&NAM\\
$T_{\rm S}$&$43\pm4.4$&SNA\\
\\
\multicolumn{3}{c}{Predictions for the values of ${\rm SN}_{\rm S}$}\\
At the epoch &Predicted value&Used value of\\
$T_{\rm M}$&$73\pm6.8$&RSA\\
$T_{\rm W}$&$78\pm6.8$&WSA\\
$T_{\rm N}$&$41\pm6.8$&NSA\\
$T_{\rm S}$&$79\pm6.8$&SAM\\
\\
\multicolumn{3}{c}{Predictions of for the values of ${\rm SN}_{\rm T}$}\\
At the epoch &Predicted value&Used value of&${\rm SN}_{\rm N} + {\rm SN}_{\rm  S}$\\
$T_{\rm M}$&$127\pm12$&RWA&$\approx 118$\\
$T_{\rm W}$&$135\pm12$&WAM&$\approx 126$\\
$T_{\rm N}$&$106\pm12$&NWA&$\approx 98$\\
$T_{\rm S}$&$131\pm12$&SWA&$\approx 122$\\
$T_{\rm M}$&$130\pm12$&RWA$^\mathrm{a}$\\
\hline
\end{tabular}
\label{table4}
}
\end{table}

\begin{SCfigure}
\setcounter{figure}{11}
\centering
\vspace{1.0cm}
\hspace{-1.0cm}
\setcounter{figure}{12}
\includegraphics[width=9.0cm]{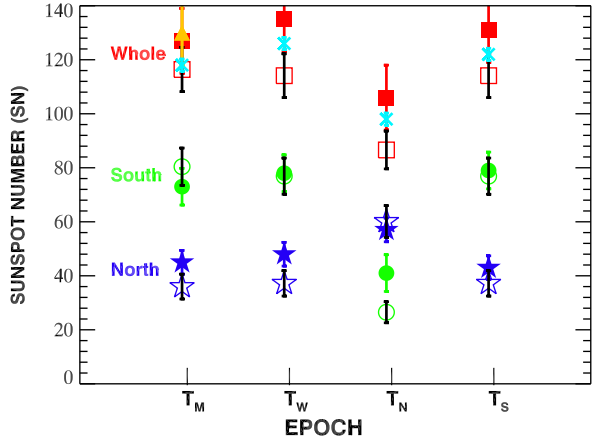}
\caption{\tiny Predicted  values of SN given in Table~4   at
   $T_{\rm M}$, $T_{\rm W}$, $T_{\rm N}$, and $T_{\rm S}$,  
 i.e. at the epochs of maxima   of 25th SN, WSGA, NSGA, and
 SSGA cycle, respectively ({\it filled symbols})   and the 
 observed values of SN  of  the corresponding 24th cycles 
({\it open symbols}) versus the maximum epochs  (the differences between 
the epochs of $X$-axis are not equal). 
 The symbols {\it square}, {\it star}, and {\it circle} represent 
 ${\rm SN}_{\rm T}$,  ${\rm SN}_{\rm N}$,  and  ${\rm SN}_{\rm S}$ 
 values, respectively (note that ${\rm SN}_{\rm T}$ at $T_{\rm M}$ is the 
 same as $R_{\rm M}$). The {\it first}, {\it second}, {\it third},
 and {\it fourth columns} contain
 the values of ${\rm SN}_{\rm T}$,  ${\rm SN}_{\rm N}$,   
and  ${\rm SN}_{\rm S}$  at the maximum epochs of  ${\rm SN}_{\rm T}$, WSGA, 
NSGA, and  SSGA Cycles~24 and 25, respectively. 
 The {\it cross} represents the value of ${\rm SN}_{\rm T}$ 
obtained by summing the  predicted values of  ${\rm SN}_{\rm N}$  and 
 ${\rm SN}_{\rm S}$.  The {\it filled triangle} 
represents the value of $R_{\rm M}$ obtained by using the value of RWA
obtained from WAM--RWA relationship shown in Figure~8}
\label{f13}
\end{SCfigure}

As shown in Figures~5\,--\,7 of \cite{jj21}, here we also 
 attempted to
 determine the linear relations between RWA and $R_{\rm M}$, RWA and RNA, 
and RWA and RSA. 
Here in the calculations of linear least-squares  fit the 
 errors in the values of both  abscissa and  ordinate are taken into account, 
whereas 
in that earlier paper, only errors in the ordinate values were considered.  
None of the corresponding linear least-squares best fits 
was found to be reasonably statistically significant, that is,
 the corresponding 
values of $\chi^2$ are found to be very large and, obviously the corresponding 
values of PROB are found to be very small ($\le$0.07).  
If we use in Equation~5 of Javaraiah (2021) the values $\approx$1317 \,msh and
$\approx$1348 \,msh obtained/predicted above for RWA (i.e. $A_{\rm W}$ in that
 paper), then we get $129\pm19$ and $131\pm19$, respectively,  for $R_{\rm M}$
 of Sunspot Cycle~25. Each of these  values of $R_{\rm M}$ of Sunspot Cycle~25
 is slightly larger than the observed value of $R_{\rm M}$ of 
Sunspot Cycle~24. As already mentioned above the
 predicted values of RWA of Solar Cycle~25 are considerably smaller than 
 the observed value of RWA of Solar Cycle~24.  However, this 
  comparison seems to be not genuine  because  the latter is an outlier
 (far from one-rms 
level)  in the $A_{\rm W}$--$R_{\rm M}$ linear relationship shown in Figure~6 
of \cite{jj21}.

We determined the best-fit linear relation
between   the 13-month smoothed monthly mean values of 
 sunspot-group area and  SN.
Figures~10, 11, and 12 show the relations between NSGA and ${\rm SN}_{\rm N}$, 
SSGA and   ${\rm SN}_{\rm S}$,  and WSGA and ${\rm SN}_{\rm T}$, 
 where the subscripts N, S, and T indicate  north,  
south, and total. All these  linear relationships 
 are statistically highly significant (the values of  PROB are high).     
By using in these relations the values predicted above for RWA, RNA, RSA,
 WNA, WSA, NAM, and SAM 
we get/predict the values of ${\rm SN}_{\rm N}$, ${\rm SN}_{\rm S}$, and
  ${\rm SN}_{\rm T}$ (it is nothing but $R_{\rm M}$ at the epoch of 
$R_{\rm M}$/RWA)  
of Sunspot Cycle~25 and the corresponding values at 
the  maxima of  25th  NSGA cycle, SSGA cycle, and WSGA cycle. 
The predicted values are also  shown in Figures~10\,--\,12 and listed in 
Table~4,  and  in Figure~13 
we compare  these predicted values of Sunspot Cycle~25 with the corresponding  
 observed values of Sunspot Cycle~24.
The  values shown in Figure~13 
 for $R_{\rm M}$ of Solar Cycle~25 are considerably larger 
than the value ($86\pm18$) predicted  earlier by \cite{jj21} by   
using the value predicted for RWA from a different method. As we can see in 
Figure~13 the  values   predicted here for  $R_{\rm M}$ of Solar Cycle~25  also 
appear to be  slightly larger  than that of Solar Cycle~24 (the corresponding 
difference is negligible if we take the uncertainty into account). 
 There is also a suggestion that the activity in the southern hemisphere 
is dominant in Solar Cycle~25, similarly to the case of sunspot-group area
 shown in Figure~9, except at the peak of NSGA, in all the remaining cases 
including the case of $R_{\rm M}$.
 The southern hemisphere's peak may  coincide with whole sphere's peak.
In fact, the overall pattern of the predicted values of Solar
 Cycle~25  closely 
resembles to that of Solar Cycle~24. Opposite to that seen in Figure~9  
for the case of sunspot-group area, in Figure~13 there is a suggestion that on 
 average Sunspot Cycle~25 would be slightly larger than Sunspot Cycle~24. 
The reason for this difference  is not clear, but   one possibility is  that the
 ratio of small to large sunspots during the maximum of Solar Cycle~25  
would be  larger than that during the  maximum of Solar Cycle~24. However,   
 the differences between    Solar Cycles 24 and 25 in sunspot-group 
area  and also between  
Sunspot Cycles~24 and 25 are statistically insignificant.      
On the other hand  properties of
 sunspot number  and  sunspots area  cycles  are not exactly the same. 
For example, the
well-known Waldmeier effect of sunspot cycles is not present in the
 cycles of sunspot-group area (\opencite{dgt08}; \opencite{jj19}).
In several solar
cycles,  there exist some differences in the  maximum epochs of sunspot-number
 and sunspot-area cycles and there are also  differences  in the relative
heights of the peaks of some  sunspot-number and sunspot-area cycles.

\begin{figure}
\setcounter{figure}{11}
\centering
\begin{subfigure}
\setcounter{figure}{13}
\includegraphics[width=5.6cm]{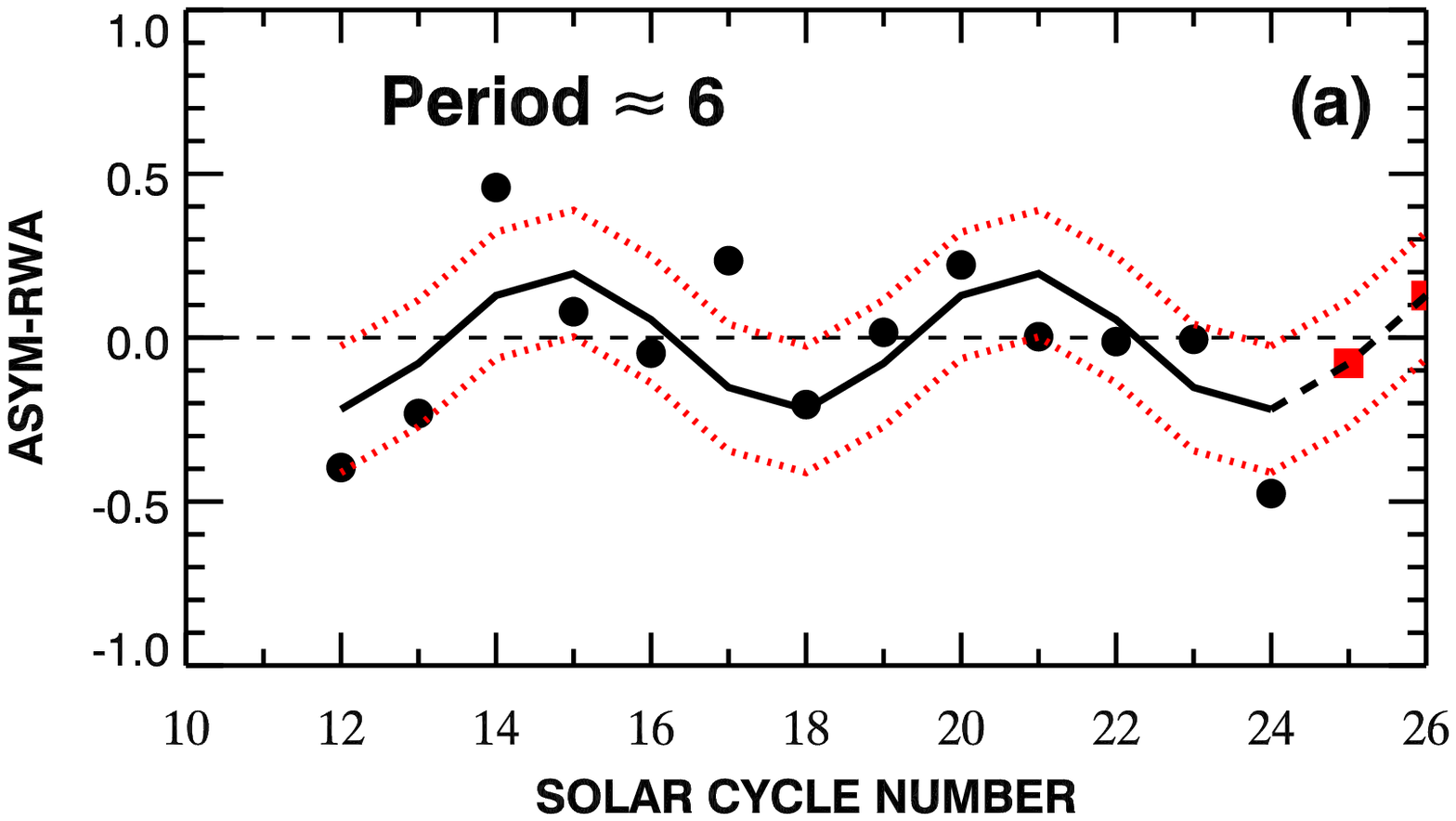}
\includegraphics[width=5.6cm]{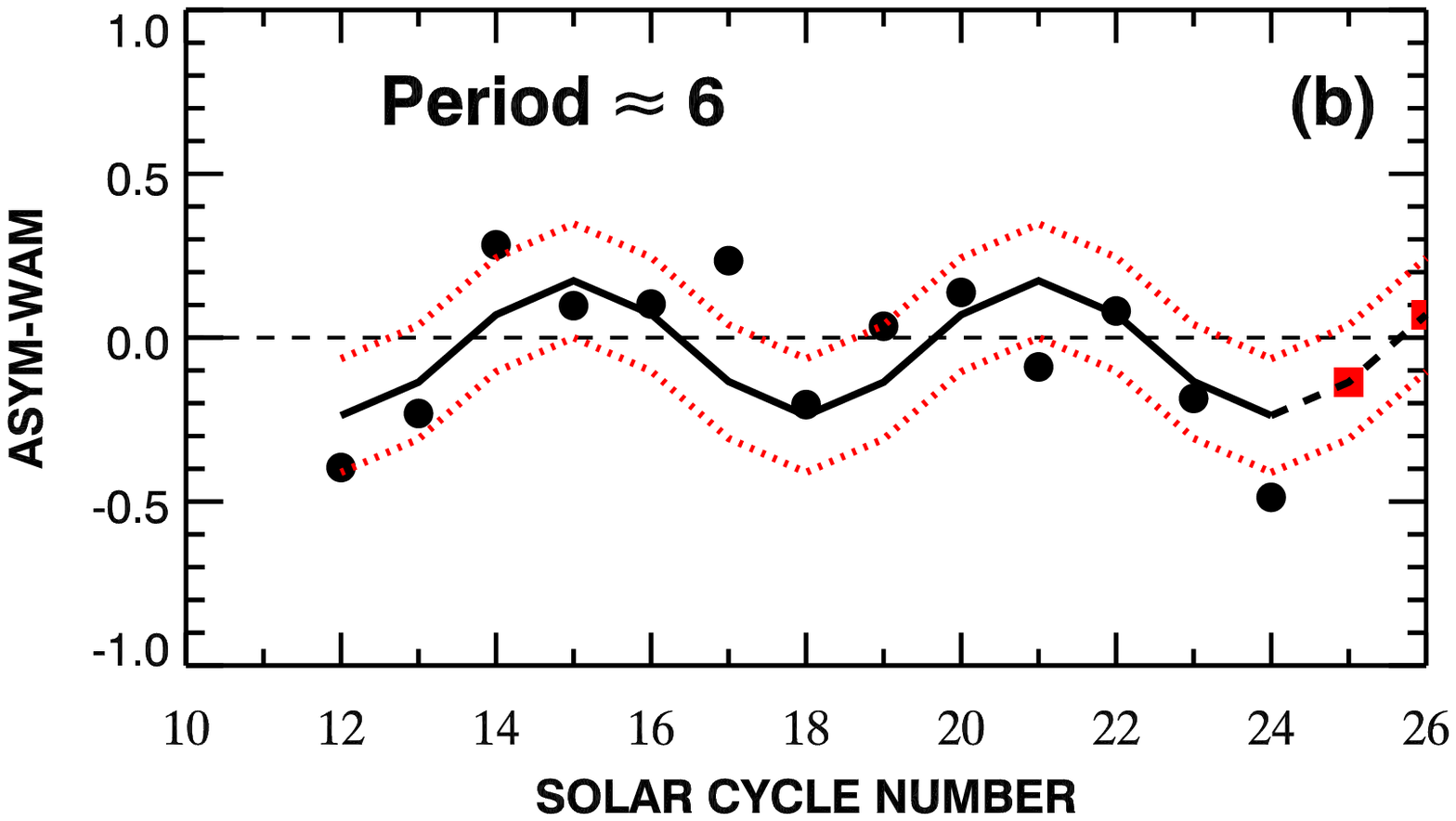}
\end{subfigure}
\begin{subfigure}
\includegraphics[width=5.6cm]{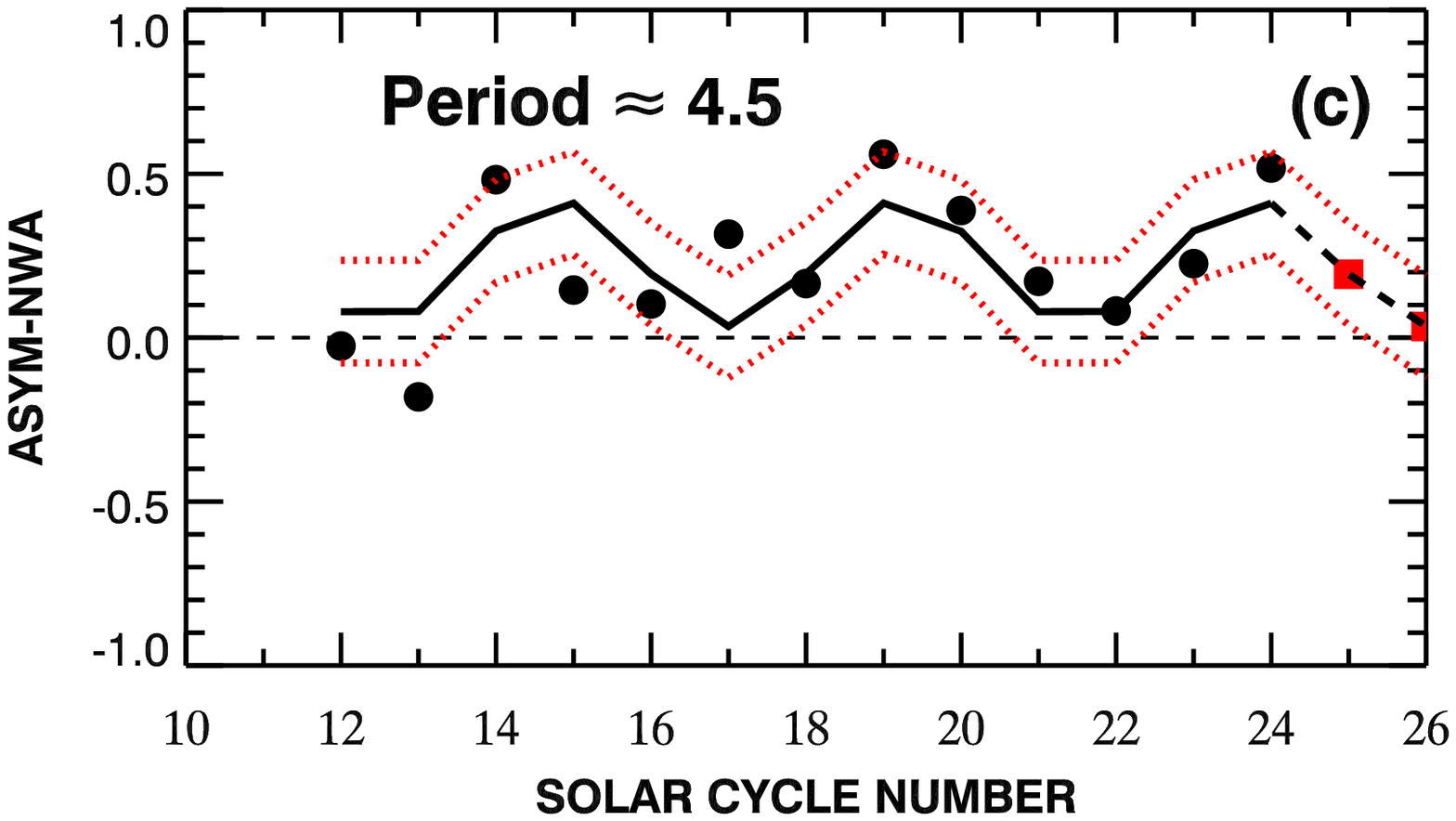}
\includegraphics[width=5.6cm]{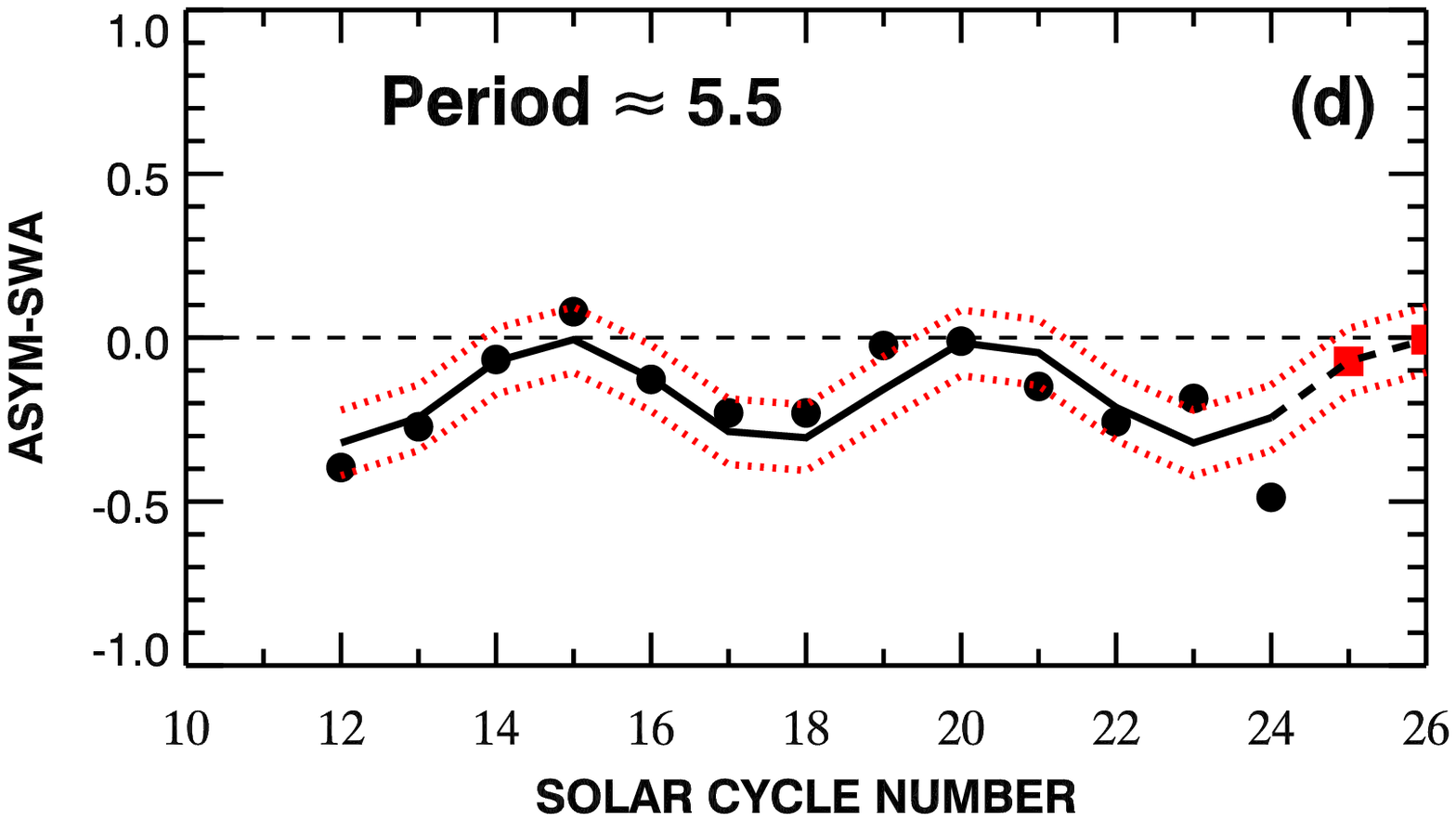}
\end{subfigure}
\caption{The {\it continuous curve} represents the best-fit cosine function
to the values   of  the relative north--south asymmetry ({\bf a}) 
in RWA, i.e. (RNA$-$RSA)/(RNA$+$RSA), ({\bf b})    in WAM i.e.
 (WNA$-$WSA)/(WNA$+$WSA),  ({\bf c})  in NWA, i.e. (NAM$-$NSA)/(NAM$+$NSA), 
and  ({\bf d}) in SWA, i.e. (SNA$-$SAM)/(SNA$+$SAM). 
 The {\it dotted curve} ({\it red}) represents the one-rms  level.
The   period (in cycles) of the best-fit cosine function  is also shown.
 The {\it dashed portion of the curve} is obtained by  extrapolation and the 
{\it filled squares} ({\it red}) represent
 the predicted value of the north--south asymmetry in the corresponding 
parameters of Solar Cycles~25 and 26.} 
\label{f14}
\end{figure}

\begin{SCfigure}
\setcounter{figure}{13}
\centering
\vspace{-0.5cm}
\hspace{-0.5cm}
\setcounter{figure}{14}
\includegraphics[width=8cm]{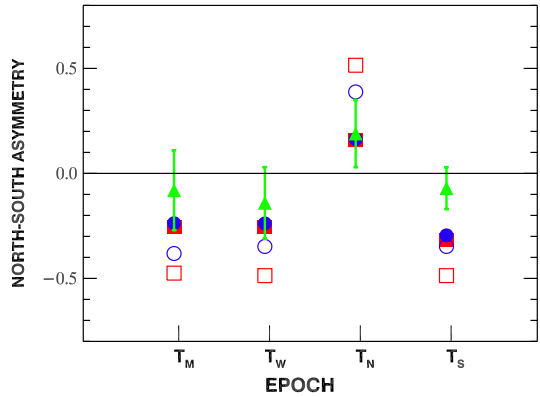}
\caption{\tiny {\it Open square} ({\it filled square}) and 
 {\it open circle} ({\it filled circle}) 
represent  the values of north--south asymmetry in the  observed (predicted)
sunspot-group area  and SN shown in Figures~9 and 13, respectively, 
   at   $T_{\rm M}$, $T_{\rm W}$, $T_{\rm N}$, and 
 $T_{\rm S}$, i.e.  at the epochs of maxima of the 24th (25th) 
  ${\rm SN}_{\rm T}$,
 WSGA, NSGA, and  SSGA cycles, respectively  (no real-time differences 
are indicated along the $X$-axis). The {\it filled triangle} 
represents the  values of the north--south asymmetry in the sunspot-group area
 at the maximum epochs of the 25th   ${\rm SN}_{\rm T}$, 
 WSGA, NSGA, and  SSGA cycles   obtained by extrapolating 
the best-fit  cosine curves shown in Figure~14.}      
\label{f15}
\end{SCfigure}

 As already mentioned above, because the cosine best-fits of most of 
the parameters are not very accurate,  although  the area--SN 
linear relationships have  small rms values 
 it may be still difficult to draw a definite conclusion 
from these predicted values 
about the  dominant hemisphere.   On the other 
hand, uncertainty in the relative north--south asymmetry of
 any two quantities is much smaller 
than those of the corresponding absolute values (\opencite{jg97}). 
Therefore we 
calculated the relative north--south asymmetry in RWA, WAM, NWA, and SWA 
 and determined the cosine fit to the relative north--south 
asymmetry in each of these quantities. It would help,  
 besides to find the long-term periodicity in north--south asymmetry 
 of each quantity, to  predict the  north--south asymmetry in the 
corresponding quantities of future cycles. 
 Figures~14a, 14b, 14c, and 14d  
show the best-fit cosine curves  of the relative north--south asymmetry 
 (RNA$-$RSA)/(RNA$+$RSA) in RWA,
 (WNA$-$WSA)/(WNA$+$WSA) in WAM,  (NAM$-$NSA)/(NAM$+$NSA) in NWA, 
and  (SNA$-$SAM)/(SNA$+$SAM) in SWA, respectively. 
The best-fit cosine curves suggest the existence of $\approx$6-cycle (66-year),
 $\approx$6-cycle (66-year), $\approx$4.5-cycle (50-year), and 
$\approx$5.5-cycle (61-year)  
 periodicities for the north--south asymmetry in RWA, WAM, NWA, and SWA, 
 i.e. at the epochs of $R_{\rm M}$, WAM, NAM, and SAM, respectively. 
We have extrapolated the 
 best-fit cosine curves and predicted the relative north--south asymmetry 
in these quantities at the maximum epochs of the  25th SN,  WSGA,
 NSGA, and SSGA cycles.  We find the values 
$-0.08\pm 0.19$, $-0.14\pm 0.17$, $0.19\pm 0.16$, and $-0.07\pm 1.0$  
for the north--south asymmetry in RWA, WAM, NWA, and SWA 
of these cycles, respectively.
Most of these  values do not significantly differ from zero.   
Figure~15 shows, besides  these values, 
  the values of  north--south  asymmetry in 
 sunspot group area and sunspot number shown in Figures~9 and 13, respectively, 
 that is,  in this figure essentially we compare the predicted values 
  of the aforementioned 25th cycle with the observed values of
 the corresponding 24th cycle. As we can see in this figure,  
 the pattern of the predicted value of 25th cycle is almost 
the same as that of the observed values of the 24th cycle,
 that is there is a suggestion that     except 
at the epoch of NAM, in all the remaining  cases,
 including at the epoch of   $R_{\rm M}$, the 
southern hemisphere is dominant during both  Solar Cycles 24 and 25.
 There is a suggestion that  overall  the asymmetry in
 Solar Cycle~25 would be smaller than that of Solar Cycle~24.

\section{Discussion and Conclusions}
In most of the solar cycles,  activity in the
 northern and southern hemispheres peaks at different times, that is  
 at least one of the hemispheric maxima frequently does not coincide with
 the  whole sphere (total)  maximum. 
Typically, one hemisphere peaks well before the other, and the maximum of the 
whole-Sun area or sunspot number can be the peak in one hemisphere, or the peak  in the other hemisphere, or something in between. 
 Predictions for each hemisphere's maximum and the corresponding north--south 
 asymmetry of a solar cycle   may  help  to  understand 
the mechanisms of the solar cycle,   the 
 solar-terrestrial relationship, and
 solar-activity influence on space weather.
Here we analysed the GPR and DPD sunspot-group data during 1874\,--\,2017 
and studied the cycle-to-cycle variations  in the values of WSGA, NSGA, and 
SSGA at the epochs of the 
maxima of Sunspot (SN) Cycles 12\,--\,24 and at the epochs of the maxima of 
  the 12th to 24th  WSGA, NSGA, and SSGA cycles. We find that,       
except in the cases of SSGA  at the peak of WSGA and NSGA at the peak of SSGA, 
in all the remaining  cases 
 the G--O rule  is violated by the solar cycle pair (22, 23).
 Using the G--O rule,  
 we obtained a value of SSGA at the maximum of WSGA Cycle~25 that is  much
 larger than the value of SSGA at the maximum of WSGA Cycle~24. Similarly, 
we  obtained  a value of NSGA at the maximum of SSGA Cycle~25 that is also 
found to be larger than the value of NSGA at the maximum of SSGA Cycle~24. 
However,  these predictions are somewhat unreliable because 
the G--O rule cannot be used to predict the amplitude of an odd-numbered 
solar cycle 
without prior knowledge about  non-violation of this rule by the
 corresponding pair of even- and odd-numbered solar cycles.  We determined the 
best-fit cosine functions to the   values of  each of the 
parameters given in Table~1. We find that there exists  a $\approx$132-year
 periodicity in all the  northern hemisphere's parameters and 
the 54\,--\,66-year periodicity in all the southern 
hemisphere's parameters.  By extrapolating the best-fit cosine curves  
 we make  cautious predictions for the values of the parameters of the
  corresponding  25th cycles.  The sums of the predicted values of the
 parameters of northern and southern hemispheres represent the 
predictions for the corresponding value of the whole sphere.  
It is found that the amplitudes of the  25th WSGA and SSGA cycles would be
to some extent  smaller than those  of the corresponding 24th cycle.
The amplitude of  25th  NSGA cycle would be almost the
same as that of the corresponding 24th cycle. The value of WSGA  at
the maximum epoch of Sunspot Cycle~25 would also be smaller than that
 at the maximum epoch of Sunspot Cycle 24,
i.e. the overall  Solar Cycle~25 in sunspot-group area would be smaller
than   Solar Cycle 24 in sunspot-group area. 
We obtained a 
good linear relationship between the 13-month smoothed monthly mean SN and 
sunspot-group area in the whole sphere and also in the northern and the  
southern hemispheres. By inputting the  predicted values of
 RWA, RNA, and RSA in  these relationships we obtained the values of SN 
at the epochs of maxima of the 25th  SN,  WSGA, NSGA, and SSGA cycles. The 
 obtained  value ($130\pm12$) of $R_{\rm M}$ of Sunspot Cycle~25 is
 found to be
 slightly larger than   that of  Sunspot Cycle~24. Except
at the maximum of NSGA Cycle~25 where the north is  found to be dominant
 in strength of activity,
 the values of SN at the maxima of  WSGA Cycle~25, SSGA Cycle~25, and
 Sunspot Cycle~25 the south is found to be dominant. 
The cosine best-fits to the values of  the corresponding north--south asymmetry 
in WSGA, NSGA, and SSGA at the maxima of the  12th to 24th  
 SN,  WSGA, NSGA, and SSGA cycles  suggest the existence of 50\,--\,66-year 
periodicities in the north--south 
asymmetry and the same aforementioned pattern of hemispheric dominance.
Overall, the present analysis 
suggests that the amplitude of Solar Cycle~25 would be  slightly 
larger than that  of Solar Cycle 24 and the activity in the southern hemisphere
 would be dominant.

All the parameters of the solar cycles of a hemisphere have the same 
 periodicity,  implying  that different parameters of the hemisphere
  are at  different phases of the same long-period cycle; 
 that is, the epochs of different
 parameters of the northern hemisphere  represent different          
phases of the same cosine wave of $\approx$132-year period
and   the different
 parameters  of the southern hemisphere represent different          
phases of the same cosine wave of 54\,--\,66-year period. 
The large difference in the periods of the cosine waves
 of northern and 
southern hemispheres' activity may be responsible for varying
 phase differences between 11-year cycles of northern and southern 
hemispheres' activity. It may be worth   finding  
why there exists a large difference in the long-term periodicities in  
northern and southern hemispheres'  activity (which is beyond the scope of 
the present analysis).

The best-fit cosine functions of sunspot-number data have already indicated 
that  Solar Cycle~25 is a reasonably small cycle and constitutes      
the minimum of the upcoming new Gleissberg 
cycle (\opencite{jbu05}; \opencite{jj17}).
A number of studies on long-term variations in solar activity 
by using different 
techniques have also predicted  Dalton-minimum-like characteristics 
around Solar Cycle~25 (\opencite{jj15}; \opencite{goa16}; \opencite{cob21}). 
 In the present analysis, we also  find the same.  However, whether 
Solar Cycle~25 will be smaller or larger than the reasonably small  
Solar Cycle~24 is yet to be established. Many authors predicted 
 the amplitude of Solar Cycle~25 by using different techniques (for 
 recent reviews see \opencite{pet20}; \opencite{nandy21}). 
Some  authors predicted a small 
amplitude, some others predicted a large amplitude for Solar Cycle~25, and
  a larger number of authors 
predicted that it would be approximately the same size as 
 Solar Cycle~24.

It is widely believed that the strength of polar fields at the minimum of an
 upcoming solar cycle is a good  physically oriented precursor to predict 
the strength of the solar cycle.  From this method and in flux-transport 
dynamo models 
it is  predicted that  Solar Cycle~25  would  have  the same size or slightly
larger than that of 
Solar Cycle 24 (\opencite{nandy21}; \opencite{kumar21}) and 
our prediction here  closely agrees with it.
As already mentioned  earlier we found a  specific 
method (\opencite{jj07}), and using  it we predicted a 
 small value for the amplitude of Solar Cycle 25,  
  strongly implying that
 Solar Cycle 25 will be considerably smaller than Solar Cycle~24 
(\opencite{jj15}; \citeyear{jj17}; \citeyear{jj21}). However,  
the predictions  made in the present analysis by using a basic  
method  suggest that  Solar Cycle~25 in   sunspot-group area would be
 slightly smaller than Solar Cycle~24 in  sunspot-group area, 
but Sunspot Cycle~25 would be slightly
 larger than Sunspot Cycle~24. A reason for this 
discrepancy in our present and earlier predictions for Sunspot Cycle~25 is
 not known to us.  There are some 
inconsistencies in   both  earlier and  present analyses. 
 Regarding the earlier method, in the case of  Solar Cycle~24 its 
second SN peak
 (coinciding with the strong southern hemisphere's peak, see Figure~1)  is
taken into account because  it is substantially larger than
its first peak (coinciding with the weak northern hemisphere peak),
 whereas in the case of most of the previous cycles the respective first peak
 is taken into account because it is larger than the corresponding
second peak. In addition, since the  uncertainties in the 
 measured areas of sunspot groups are not known,  in the earlier analyses  
  the linear least-squares fit calculations were not 
included the uncertainties in the area-sums 
  (sums  of the areas of sunspot groups; see Section 1). 
 Here we repeated the   
earlier calculations by taking 10\,\% of the area-sums for their uncertainties 
 (note that $R_{\rm M}$ values have 5\,\%\,--\,10\,\% uncertainties). None of
 the best-fit linear relations, namely between the area-sum and $R_{\rm M}$ and 
area-sum and RWA, etc. are found statistically significant  
(PROB values  are found considerably small). However,  the  10\,\%
 uncertainties 
in the area-sums seem to be underestimated.  To get a reasonably 
large PROB it seems necessary to use 20\,--\,25\,\% of the area-sums
 for their 
uncertainties, which may be too  unrealistic.  In the present analysis the 
results of 
 best-fit cosine functions are suggestive rather than compelling
 (best-fit cosine functions have to some extent  large $\chi^2$-values). 
The predictions  made from our earlier method are  based on 
the existence of a strong connection (high correlation)  between 
consecutive cycles. The values predicted  here from the long-term
 cyclic trends in the data might be to some extent overestimated.
Further investigations are needed to   conclude 
whether our  present or earlier predictions  will be  right or wrong.
 On the other hand the value predicted in the present analysis 
 for the  amplitude  of Solar Cycle~25
 is   only slightly larger than that of the reasonable small Solar Cycle~24
 and considerably smaller than the average amplitude 178.7 of solar cycles
(\opencite{pesnell18}). Hence in the present analysis we  confirmed that
the beginning  of the upcoming Gleissberg cycle 
  would take place    around  Solar Cycle~25 (see the cosine curve 
in Figure~7). 

 Our present predictions (see Table~4)  for Solar Cycle~25 are somewhat 
comparable to  the following  predictions  made by different authors 
 using different techniques. 
\cite{rigozo11} by extrapolating  the sunspot-number time-series
 spectral components estimated a maximum sunspot number
 of 132.1 for Solar Cycle 25.
\cite{cameron16} by using a flux-transport dynamo model  simulated a
value of dipole moment around 2020, i.e. around the beginning of 
Solar Cycle~25, and suggested that Solar Cycle~25 will be of
moderate amplitude, not much higher than that of Solar Cycle~24.
\cite{ps18}  applied a precursor method by
using the SODA Index, which is determined by polar magnetic fields
and the spectral index,  predicted a large amplitude ($135 \pm 25$) 
 for Solar Cycle~25.
\cite{jiang18} used a solar surface flux-transport  model
 and predicted that the peak sunspot number of Solar Cycle~25 
is 125$\pm$ 32.
\cite{pet18} using  coronal green-line data as a precursor 
predicted 130 for the amplitude of Solar Cycle~25.
\cite{bn18} using magnetic-field evolution models for the Sun's surface
and interior predicted that Solar Cycle 25 will be similar or
slightly stronger than  Solar Cycle~24. 
\cite{du20} using  the rate of
decrease in the smoothed monthly mean sunspot number  over the 
final three years of  Solar Cycle~24 predicted $130.0 \pm 31.9$ for 
 the amplitude of Solar Cycle~25. 
\cite{kumar21} used  
the polar-precursor method and predicted $126\pm3$ for the
 amplitude of Solar Cycle~25.  
\cite{hsa22} using   data for sunspot  numbers  and 
geomagnetic indices (aa/Ap)  and  their 
 heuristic methodology  inferred that  Solar Cycle~25 
may be as active as Solar Cycle~24 ruling out a Dalton-like minimum
 in the 21st century.
\cite{uh18} using an  advective flux transport  code  found  evidence
that during Solar Cycle~25 the southern hemisphere will be  more active than
the northern hemisphere. 
\cite{gopal18} used a
precursor method (microwave-imaging observations) and predicted that
the sunspot-number peaks of southern and
northern hemispheres during Sunspot Cycle 25 will be 89 and 59, respectively.
These  indicate a considerable north--south asymmetry corresponding to the
 amplitude of Solar  Cycle 25 and that activity would be dominant in the 
southern hemisphere.
 \cite{werner20} using 
 auto-regressive models of different order predicted 117 for 
the amplitude  of Solar Cycle~25 and that
 solar activity would be dominant  in the southern hemisphere. 
\cite{pish21} using the absolute values of the mean polar magnetic fields 
during the two-year interval just before the cycle minimum in northern and 
southern hemispheres as precursors predicted $66\pm17$ and $83\pm21$ for 
northern and southern hemispheres' amplitudes of Solar Cycle~25.
 \cite{lcl19} from a data-driven solar cycle model predicted 
 a low  amplitude for Solar Cycle~25 and  
 the northern hemisphere about 20\,\% more   than 
the south southern hemisphere. 
Our earlier analysis (\opencite{jj21})  also suggested that at the
epoch of maximum of  the Solar Cycle~25 the activity will be about 20\% more
in  northern hemisphere than that in  southern hemisphere, whereas our present
analysis suggests that the southern hemisphere would be dominant.
According to the international
panel to forecast Solar Cycle 25, the amplitude  of this cycle  would be 
115, which is almost the same as the 
amplitude of Solar Cycle 24 (see \textsf{www.swpc.noaa.gov/news}).

\begin{acks}
 The author thanks the anonymous reviewer for useful
 comments and suggestions.
The author acknowledges the work of all the
 people who contribute to and maintain the GPR and DPD  Sunspot databases.
The sunspot-number data are provided by WDC-SILSO, Royal Observatory of
Belgium, Brussels.
\end{acks}
 \begin{ethics}
 \begin{conflict}
The author declares that he has no conflicts of interest.
 \end{conflict}
 \end{ethics}

\begin{dataavailability}
 All data generated or analysed during this study are included in this article.
\end{dataavailability}
 \bibliographystyle{spr-mp-sola}
 \bibliography{jaja.bib}  

\end{article}
\end{document}